\documentclass[10pt,journal]{IEEEtran}

\usepackage{xcolor}
\usepackage{cite}
\usepackage{amsmath}
\usepackage{amssymb}
\usepackage{epsfig}
\usepackage{multirow}
\usepackage{xfrac}
\usepackage{subfigure}
\usepackage{multirow}
\usepackage{booktabs}
\usepackage{pstricks}
\usepackage{pst-node}

\newtheorem{theorem}{Theorem}[section]

\allowdisplaybreaks

\ifCLASSOPTIONcompsoc
  \usepackage[nocompress]{cite}
\else
  \usepackage{cite}
\fi

\ifCLASSINFOpdf

\else

\fi

\usepackage{latexsym}

\begin{document}

\title{
  \textcolor{black}{
    Heterogeneous Networked Data Recovery from Compressive Measurements
   Using a Copula Prior
 }
}

\author{Nikos~Deligiannis,
        Jo\~{a}o~F.~C.~Mota,
        Evangelos~Zimos,
        and~Miguel~R.~D.~Rodrigues,
\IEEEcompsocitemizethanks{
This work has been presented in part at the Data Compression Conference 2016~\cite{zimos2016bayesian} and the International Conference on Telecommunications 2016~\cite{zimos2016internet}.}
\IEEEcompsocitemizethanks{
\IEEEcompsocthanksitem N.~Deligiannis and E. Zimos are with the Department of Electronics and Informatics, Vrije Universiteit Brussel, Pleinlaan 2, 1050 Brussels, Belgium and with  imec, Kapeldreef 75, B3001, Leuven, Belgium. E-mail: ndeligia@etrovub.be, ezimos@etrovub.be.
\IEEEcompsocthanksitem J.~F.~C.~Mota is with the Institute of Sensors, Signals and Systems, Heriot-Watt University, Edinburgh EH14 4AS, UK. Email: j.mota@hw.ac.uk.
\IEEEcompsocthanksitem M.~R.~D.~Rodrigues is with the Electronic and Electrical Engineering Department, University College London, Torrington Place, London WC1E 7JE, UK. 
E-mail: m.rodrigues@ucl.ac.uk.}
}

\markboth{}%
{Shell \MakeLowercase{\textit{et al.}}: Bare Advanced Demo of IEEEtran.cls for IEEE Computer Society Journals}
%

\definecolor{red}{RGB}{153,0,0}

\newcommand{\bm}[1]{\boldsymbol{#1}} 

\IEEEtitleabstractindextext{%
\begin{abstract}
Large-scale  data collection by means of wireless sensor network and internet-of-things technology
poses various challenges in view of the limitations in transmission, computation, and energy
resources of the  associated wireless devices. 
\textcolor{black}{
  Compressive data gathering  based on
  compressed sensing has been proven a well-suited solution to the problem. Existing designs exploit
  the spatiotemporal correlations among data collected by a specific sensing modality. However, many
  applications, such as environmental monitoring, involve collecting heterogeneous data that are
  intrinsically correlated.  In this study, we propose to leverage the correlation from multiple
  heterogeneous signals when recovering the data from compressive measurements.
}
To this end, we propose a novel recovery algorithm---built upon belief-propagation principles---that
leverages correlated information from multiple heterogeneous signals. To efficiently capture the
statistical dependencies among diverse sensor data, the proposed algorithm uses the statistical model
of copula functions. Experiments with heterogeneous air-pollution sensor measurements  show that the
proposed design provides significant performance improvements against
state-of-the-art compressive data gathering and recovery schemes that use classical compressed
sensing, compressed sensing with side information, and distributed compressed sensing.
\end{abstract}

\begin{IEEEkeywords}
Compressed sensing, side information,
 copula functions, air-pollution monitoring, wireless sensor networks.
\end{IEEEkeywords}}

\maketitle

\IEEEdisplaynontitleabstractindextext

%
\IEEEpeerreviewmaketitle

\ifCLASSOPTIONcompsoc
\IEEEraisesectionheading{\section{Introduction}\label{sec:introduction}}
\else
\section{Introduction}
\label{sec:introduction}
\fi

\IEEEPARstart{T}{he} emerging paradigm of smart cities has triggered the development of new application domains, such as environmental monitoring and smart mobility.
These applications typically involve large-scale wireless sensor networks (WSNs) and internet-of-things (IoT) devices collecting and communicating massive amounts of environmental data, related to air pollution, temperature, and humidity. \textcolor{black}{An air-pollution monitoring system\footnote{\textcolor{black}{One can visit the websites of the European Environment Agency (EEA)  (http://www.eea.europa.eu/themes/air/air-quality) and the USA Environmental Protection Agency (https://www.epa.gov/aqs).}}, for example, involves wireless devices spread in an urban area communicating measurements on several
air pollutants, including carbon monoxide (CO), nitrogen dioxide ($\text{NO}_2$), ozone
($\text{O}_3$), and sulfur dioxide ($\text{SO}_2$).} Such data types have very different ranges and marginal
statistics, but are intrinsically correlated. 

\textcolor{black}{This work shows how to effectively leverage the dependencies among diverse (alias,
  heterogeneous) data types in order to significantly reduce data-rates in the network.} This
reduction translates into power savings at the wireless nodes or IoT devices, which operate under
austere limitations in energy resources. Efficient designs should, nevertheless, exploit  intra- and
inter-data  dependencies at the decoder so as to conserve  the computational effort at the wireless
sensors and to diminish energy-demanding inter-sensor communication. Moreover, in order to safeguard
power savings, devices should communicate over small distances through multi-hop wireless
transmissions \cite{akyildiz2010wireless}, namely, from neighbor to neighbor, rather than directly to a sink.   Finally, as information is sent over error-prone wireless channels, data collection and recovery schemes should provide for robustness against communication noise.


\subsection{Prior Work}

\textcolor{black}{Related studies  on the problem of data collection and recovery for WSNs proposed to
  reduce  data  rates  by grouping nodes with  correlated readings into
  clusters~\cite{liu2007energy,yoon2007clustered} or by allowing a small subset of nodes to transmit
  data  carrying most of the information in the network~\cite{gupta2008efficient}.}  Alternative
studies focused on conventional data compression techniques involving differential pulse-code
modulation (DPCM) followed by entropy encoding \cite{vecchio2014,sacaleanu2012compression}. Other
solutions considered collaborative wavelet transform coding \cite{crovella2003graph}
\textcolor{black}{or offered a flexible selection between a distributed  wavelet transform  and a
  distributed prediction based scheme~\cite{ciancio2006energy,acimovic2005adaptive}}.  These
techniques, however, require additional inter-sensor communication, increasing the  transmission of
overhead  information over the network. 

An alternative strategy adheres to distributed source coding (DSC)~\cite{xiong2004distributed}, a paradigm that leverages inter-sensor (spatial) data correlation   via joint decoding. \textcolor{black}{DSC is a promising technique for WSNs as it shifts the computational burden towards the sink node and delivers  code constructs that are robust against communication errors~\cite{xiong2004distributed}. However, extending DSC to the multiterminal case is known to be a challenging problem in practice~\cite{stankovic2006code,deligiannis2015distributed,chen2013compression}.}

Compressed sensing (CS)~\cite{Donoho06compressed,cande2008introduction}   addresses the problem of data aggregation  in  WSNs by enabling data   to be recovered from a small set of linear measurements~\cite{haupt2008compressed}. \textcolor{black}{CS involves solving an inverse problem at the decoder,
for which several algorithms have been proposed, including orthogonal matching pursuit (OMP) \cite{tropp2007signal}, iterative thresholding~\cite{blumensath2009iterative},
 belief propagation (BP)~\cite{baron2010bayesian}, and approximate message passing (AMP)~\cite{donoho2009message}.}

\textcolor{black}{Considering a single-hop network, Haupt \textit{et al.} \cite{haupt2006signal} proposed CS-based data aggregation through synchronized amplitude-modulated transmissions of   randomized sensor readings.} Alternatively, Duarte \textit{et al.}~\cite{duarte2005distributed} proposed distributed compressed sensing (DCS), where random measurements are transmitted from each sensor and the data are jointly recovered at the sink by leveraging the  spatiotemporal correlations. Furthermore, the authors of~\cite{masiero2009data,quer2012sensing} proposed a CS-based data aggregation  method that used principal component analysis (PCA) to capture the spatiotemporal correlations in the data. 

\textcolor{black}{Assuming multi-hop transmission, Luo \textit{et al.}~\cite{luo2010efficient} proposed a compressive data gathering method that alleviated the need for centralized control and complicated routing. They also presented measurement designs that limit the communication cost without jeopardising the data recovery performance. As an alternative solution, Lee \textit{et al.}~\cite{lee2009spatially} proposed spatially-localized projection design by clustering  neighboring nodes.}    


\subsection{Contributions}
Prior studies on networked data aggregation via  (distributed) compressed sensing~\cite{duarte2005distributed,masiero2009data,quer2012sensing,luo2010efficient} considered  \textit{homogeneous} data sources, namely, they proposed to leverage the spatiotemporal correlations within  signals of the same type. Many applications, however, involve  sensors of \textit{heterogeneous} modalities measuring diverse yet correlated data (e.g., various air pollutants, temperature, or humidity). 
\textcolor{black}{
  In this work, we propose a novel compressive data reconstruction method that  exploits both intra-
  and inter-source dependencies, leading to significant performance improvements. Our specific
  contributions are as follows:
}
\begin{itemize} 

  \item 
    \textcolor{black}{
    We propose \textcolor{black}{a new heterogeneous networked data recovery method}, which builds
    upon the concept of Bayesian CS with belief propagation \cite{baron2010bayesian}. Our algorithm
    advances over  this concept by incorporating multiple  side-information\ signals, gleaned from
    heterogeneous correlated sources.  This is in contrast to previous
    studies~\cite{mota2014compressed,mota2016ref,mota2014glob,renna2016classification}, which consider
    signal recovery aided by a single side information signal.
    }

  \item 
    \textcolor{black}{
    Previous CS approaches describe the dependency among \textit{homogeneous} sensor readings using
    the sparse common component plus innovations model \cite{duarte2005distributed}; simple additive
    models \cite{liu2012noise}; or joint  Gaussian mixture models~\cite{renna2016classification}.
    Unlike these studies, we model the  dependency among \textit{heterogeneous} data sources  using
    copula functions~\cite{sklar1959fonctions,nelsen2006introduction} and  we  explore copula-based
    graphical models---based on belief propagation \cite{mackay2003information}---for data recovery.
    Copula functions model the marginal distributions and the dependence structure among the data
    separately; as such, they capture complex dependencies among diverse data more accurately than
    existing  approaches. 
    }


  \item 
    \textcolor{black}{
    Experimentation \textcolor{black}{using synthetic data as well as diverse air-pollution sensor measurements} from  the USA Environmental
    Protection Agency \cite{epa} shows that, for a given data rate, the proposed method reduces the
    reconstruction error of the recovered data with respect to classical CS~\cite{luo2010efficient}, CS
    with side information~\cite{mota2014compressed}, and DCS~\cite{duarte2005distributed} based
    methods. Alternately, for a given reconstruction quality, the method offers significant rate
    savings, thereby resulting in less network traffic and \textcolor{black}{reduced energy consumption at the wireless devices}. Furthermore, the
    proposed design offers increased robustness against \textcolor{black}{imperfections in the
      communication medium} compared to the classical CS~\cite{luo2010efficient} and
    DCS~\cite{duarte2005distributed} based methods.
    }

\end{itemize}

\subsection{Outline} 

The paper continues  as follows: Section \ref{sec:Background} gives the  background of the work and \textcolor{black}{Section \ref{sec:PropDataRec} details  the proposed data  recovery method.} Section \ref{sec:StatModel} describes the copula-based statistical model for expressing the dependencies among diverse data types, whereas Section \ref{sec:prop_recovery} elaborates on the proposed belief-propagation algorithm. Experimental results are provided in Section \ref{sec:Experiments}, whereas Section \ref{sec:conclusion} concludes the work.   

\section{Background}
\label{sec:Background}
\subsection{Compressed Sensing}

Compressed Sensing (CS) builds upon the fact that many signals $\bm{x} \in
\mathbb{R}^n$ have sparse representations, i.e., they can be written as
$\bm{x}=\bm{\Psi}\bm{s}$, where $\bm{\Psi}\in\mathbb{R}^{n\times n_0}$ is a
dictionary matrix, and $\bm{s}\in \mathbb{R}^{n_0}$ is a $k$-sparse vector (it has
at most $k$ nonzero entries). Suppose we observe $m \ll n$ linear measurements from $\bm{x}$:
$\bm{y}=\bm{\Phi}\bm{x}=\bm{A}\bm{s}$, where $\bm{\Phi}\in \mathbb{R}^{m\times n}$
is a sensing (or encoding) matrix, and $\bm{A}:=\bm{\Phi}\bm{\Psi}$. CS theory
states that if $\bm{A}$ satisfies the mutual coherence property \cite{donoho2001uncertainty}, the
Restricted Isometry Property \cite{candes2005decoding}, or the Null Space Property
\cite{chandrasekaran2012convex}, then $\bm{s}$ (and thus $\bm{x}$) can be recovered by
solving
\begin{align}
\label{eq:BP}
\hat{\bm{s}}&=\arg\min_{\bm{s}}\|\bm{s}\|_1\ \ \text{s.t.}\
\ \bm{y}=\bm{A}\bm{s}. 
\end{align}
In particular, $\bm{s}$ is the only solution to \eqref{eq:BP} whenever the number of measurements $m$ is sufficiently large.
When the measurements are noisy, i.e., $\bm{y}=\bm{A}\bm{s}+\bm{z}$, where
$\bm{z} \in \mathbb{R}^m$ represents additive noise, $\bm{s}$ can be recovered by solving
instead
\begin{align}
\hat{\bm{s}}=\arg\min_{\bm{s}}\frac{1}{2}\|\bm{y}- 
\bm{A}\bm{s}\|_2^2+\kappa\|\bm{s}\|_1,
\end{align} 
where $\kappa > 0$ controls the trade-off between sparsity and reconstruction fidelity. Instead of assuming that~$\bm{s}$ is strictly sparse (i.e., $\|\bm{s}\|_0=k$), several works~\cite{baron2010bayesian} (including this one) focus on  compressible signals, i.e., signals whose coefficients decay exponentially, when sorted in order of decreasing magnitude.

\subsection{Compressed Sensing with Side Information}

CS can be modified to leverage a signal correlated to the signal of interest, called side
information, which is provided \textit{a priori} to the decoder, in order to aid
reconstruction~\cite{mota2014compressed,mota2016ref,mota2014glob,renna2016classification,mota2015dynamic}.
In CS with side information, the decoder aims to reconstruct $\bm{x}$
from the measurements $\bm{y}$, the  matrix $\bm{A}$, and a side information vector
$\bm{w}$ that is correlated with $\bm{s}$. The work in
\cite{mota2014compressed,mota2016ref,mota2014glob}  provides guarantees for a particular way of
integrating side information into CS. In particular, one adds to the objective of~\eqref{eq:BP} the
$\ell_1$-norm of the difference between the optimization variable $\bm{s}$ and the side
information $\bm{w}$, yielding the $\ell_1$-$\ell_1$ minimization problem:  
\begin{align}
\label{eq:L1L1minDefinition}
\bm{\hat{s}}=\arg&\min_{\bm{s}}\|\bm{s}\|_1
+\|\bm{s}-\bm{w}\|_1\: \: \text{s.t.}\
\ \bm{y}=\bm{A}\bm{s}. 
\end{align}
Other studies considered  prior information in the form of knowledge about the sparsity structure of  $\bm{s}$ \cite{vaswani2010modified,scarlett2013compressed,khajehnejad2009weighted,oymak2012recovery}
and derived sufficient conditions for exact reconstruction \cite{vaswani2010modified}. The authors of
\cite{trocan2010disparity} proposed to recover the difference between the signal of interest and the
side information, which was assumed to be sparser than the signal itself.

\subsection{Distributed Compressed Sensing} 

DCS~\cite{duarte2005distributed} assumes a joint sparsity model to describe the spatiotemporal
dependencies among~$\zeta$ homogeneous signals. \textcolor{black}{The sensor signals $\bm{x}_j\in
\mathbb{R}^n,j \in \{1,2,\dots,\zeta\}$, are assumed to have a representation $\bm{x}_j =
\bm\Psi(\bm{s}_c+ \bm{s}_j)$, where  $\bm{s}_{c}\in \mathbb{R}^n$ is a sparse
component common  to all signals, $\bm{s}_j\in \mathbb{R}^n$ is a sparse innovation component
unique to each signal, and $\bm\Psi \in \mathbb{R}^{n\times n}$ is the sparsifying basis.
Each sensor $j\in\{1,2,\dots,\zeta\}$ independently encodes the measured signal by projecting it onto
a sensing matrix $\bm{\Phi}_j$ and transmits the low-dimensional measurements $\bm{y}_j =
\bm{\Phi}_j \bm{x}_j$ to the sink. The sink, in turn, jointly reconstructs the signals by solving:}
\begin{align}
\nonumber
\bm{\hat{s}_{\text{all}}}=\arg\min_{\bm{s_{\text{all}}}}\:\|\bm{s}_c\|_1+\sum_{j=1}^\zeta\omega_j\|\bm{s}_j\|_1
\quad\text{s.t.}\
\ \bm{y_{\text{all}}}=\bm{A_{\text{all}}}\bm{s}_{\text{all}},
\end{align}
where $\omega_1, \ldots, \omega_\zeta>0$,  
$
\bm{y_{\text{all}}}
=
\begin{bmatrix}
\bm{y}_1^T &  \cdots & \bm{y}_\zeta^T
\end{bmatrix}^T
$
contains the measurements from all the sensors, and
$
\bm{s_{\text{all}}}
=
\begin{bmatrix}
  \bm{s}_c^T & \bm{s}_1^T & \cdots & \bm{s}_{\zeta}^T
\end{bmatrix}^T
$
the vector to be recovered, contains the common and all the innovation components. Also,
$$
\bm{A}_{\text{all}}=
\begin{bmatrix}
\bm{A}_1&\bm{A}_1&\bm{0}&\bm{0}  & \cdots & \bm{0}  
   \\
\bm{A}_2&\bm{0}&\bm{A}_2&\bm{0}  & \cdots & \bm{0}  
   \\
\vdots & \vdots  & \vdots    & \vdots                    & \ddots     & \vdots \\
\bm{A}_\zeta&\bm{0}&\bm{0}&\bm{0}    & \cdots & \bm{A}_\zeta
\end{bmatrix},
$$
\textcolor{black}{where $\bm{A}_j=\bm\Phi_j\bm{\Psi}$ is associated to sensor
$j\in\{1,2,\dots,\zeta\}$. Note that the $j$-th block equation of
$\bm{y}_{\text{all}} = \bm{A}_{\text{ext}}\bm{s}_{\text{all}}$ corresponds to the
measurements of sensor $j$: $\bm{y}_j = \bm{A}_j(\bm{s}_c + \bm{s}_j)$}.

\begin{figure}[t]
\centering
  \includegraphics[scale=0.3]{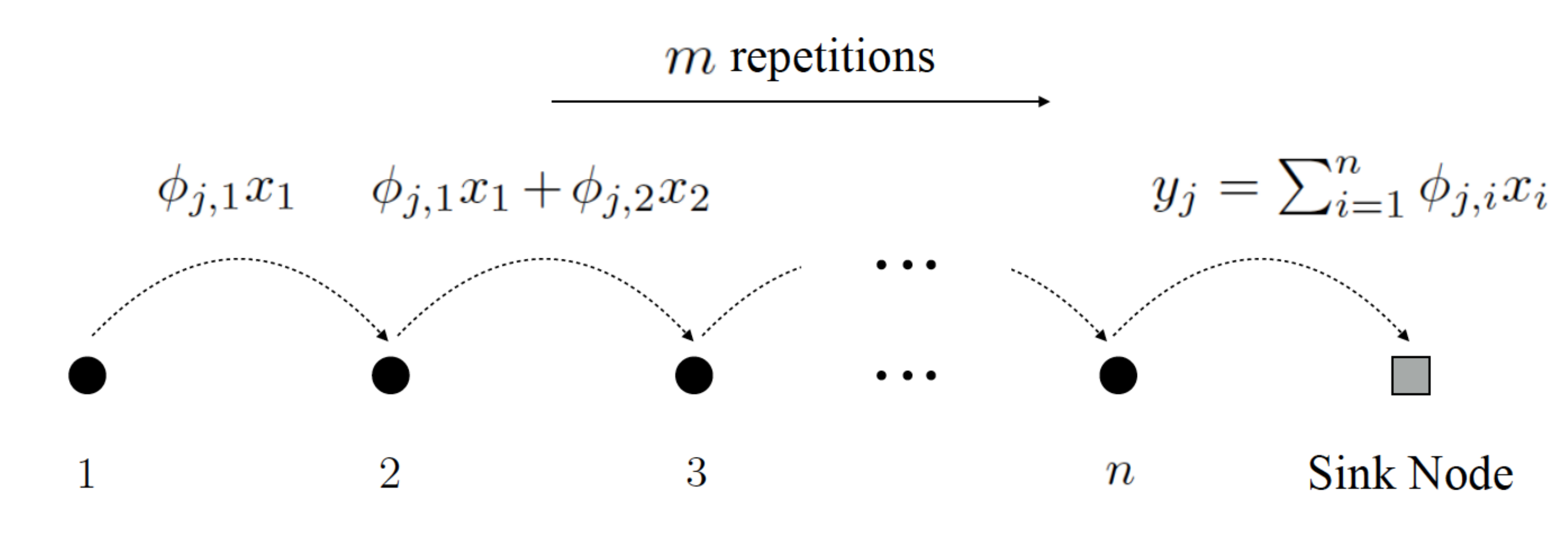} 
  \caption{Multi-hop transmission in a large-scale WSN  using CS~\cite{luo2010efficient}.}
  \label{fig:Multihop_scheme}
\end{figure}             

\subsection{Compressive Data Gathering for WSNs}
\label{sec:SoTA_scheme} 
The compressive data gathering approach in~\cite{haupt2008compressed,luo2010efficient} adheres to a
multi-hop communication scenario in which each node relays a  weighted sum of sensor readings to a
neighboring node. Specifically, consider a network of $n$ nodes and let $x_{i}\in\mathbb{R}$ denote a
scalar reading of node $i\in\{1, 2, \ldots, n\}$. \textcolor{black}{As shown in
  Fig.~\ref{fig:Multihop_scheme}, node~$1$ generates a pseudorandom number $\phi_{j,1}$ --- using its network
address as the seed of a pseudorandom number generator --- and transmits the value  $\phi_{j,1}x_{1}$
to node~$2$.} Subsequently, node $2$   generates $\phi_{j,2}$, computes the weighted sum
$\phi_{j,1}x_1+\phi_{j,2}x_2$ and sends it to node~$3$. In sum, node $k$  generates  $\phi_{j,k}$,
computes the value $\phi_{j,k}x_k$, adds it to the sum of the previous relayed values, and sends
$\sum_{i=1}^{k}\phi_{j,i}x_{i}$ to node~$k+1$. The sink node thus receives 
$y_{j}=\sum_{i=1}^{n}\phi_{j,i}x_{i}$. After repeating the procedure~$m$ times, for $j=1,\dots,m$, the 
sink obtains
\begin{equation}
\label{eq:luoDataGathering}
  \bm{y}
  =
  \begin{bmatrix}
    \bm{\phi}_1 & \cdots & \bm{\phi}_i & \cdots & \bm{\phi}_n
  \end{bmatrix}
  \bm{x}
  =
  \bm{\Phi}
  \bm{x}\,,
\end{equation} 
where $\bm{y}=(y_1,\ldots, y_j, \ldots, y_m)$ is the vector of measurements, $\bm\phi_i=(\phi_{1,i},
\ldots, \phi_{j,i}, \ldots, \phi_{m,i})$ is the column vector of pseudorandom numbers generated by
node $i$, and $\bm{x}=(x_1, \ldots, x_i,\ldots, x_n)$ is the vector of the node readings. Given the
seed value and the addresses of the nodes, the sink can replicate~$\bm{\Phi}$ and recover the data
$\bm{x}$ using standard CS recovery
algorithms~\cite{tropp2007signal,donoho2009message,baron2010bayesian}.
\textcolor{black}{
The study in~\cite{luo2010efficient} modified the sensing matrix in~\eqref{eq:luoDataGathering}
as~$\bm{\Phi}' = \begin{bmatrix}\bm{I} & \bm{R}\end{bmatrix}$, where $\bm{I}$ is the $m\times m$
identity matrix and~$\bm{R}\in\mathbb{R}^{m\times(n-m)}$ is a pseudorandom Gaussian matrix. This
means that the first~$m$ nodes transmit their original readings directly to node~$m+1$, which leads
to a reduced number of transmissions in the network.
} 

\textcolor{black}{
Alternatively, in the approach of~\cite{quer2012sensing}, each node~$i$ transmits with
probability~$p_i$  its reading directly to the sink. In this way, the sink collects measurements
$\bm{y}=\bm\Phi \bm{x}$, where $\bm\Phi$ is a very sparse binary matrix with one element equal to 1
per row and at most one element equal to 1 per column, while all the other elements are zero. The
sink then solves~\eqref{eq:BP} to recover the readings from all the nodes in the network.
}


\begin{figure}[t] 
\centering
\includegraphics[scale=0.29]{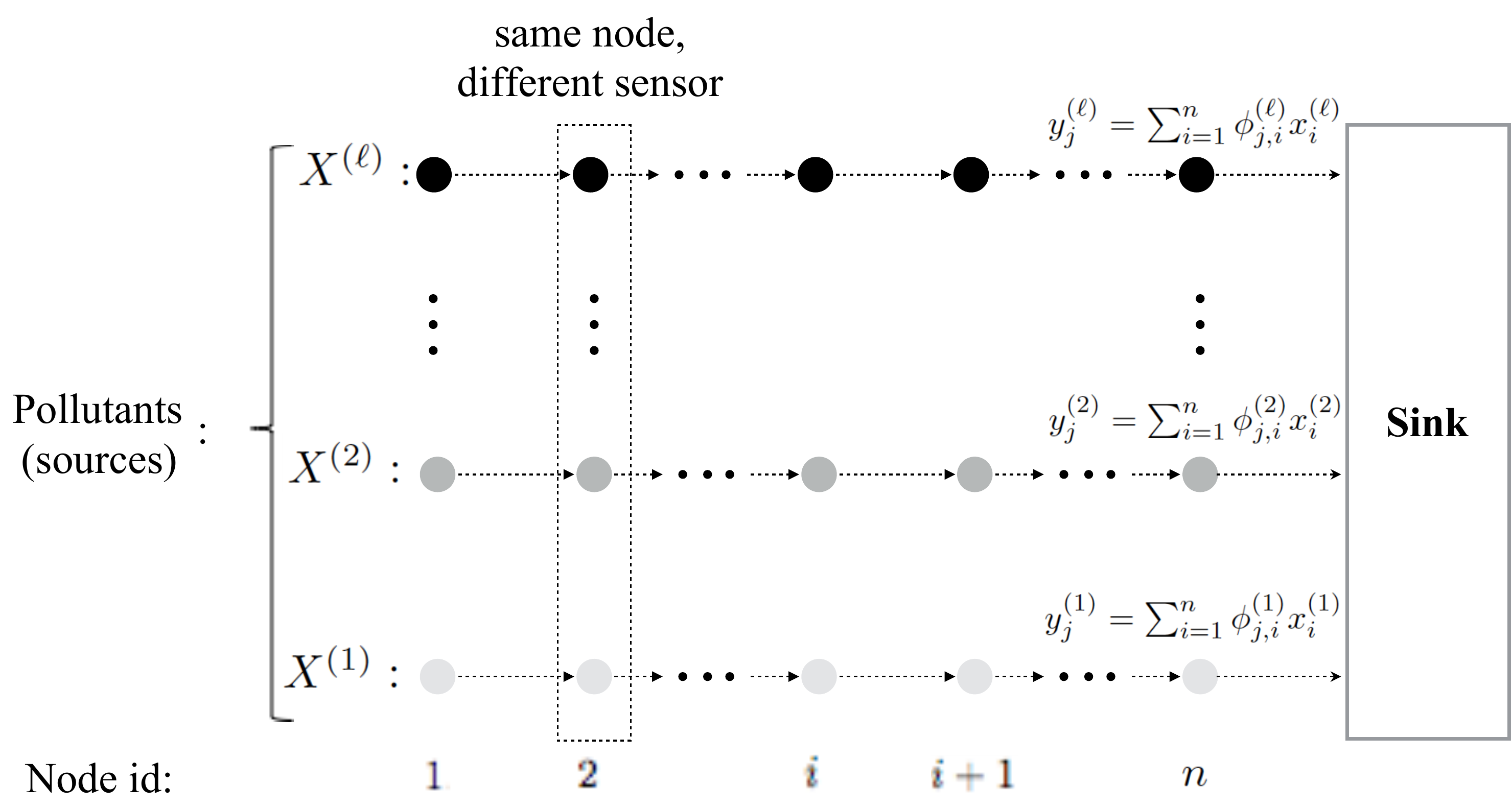}   
\caption{Extension of the data gathering scheme of~\cite{luo2010efficient} to support the collection of diverse sensor data. For each source $X^{(l)}$, $l=1,2,\dots,\ell$, the multi-hop transmission among the nodes takes place for $m^{(l)}$ repetitions until the measurements vector $\boldsymbol{y}^{(l)}$
is formed at the sink.}
\label{fig:ourEncoder}
\end{figure}

\begin{figure}[t]
\centering
\includegraphics[scale=0.45]{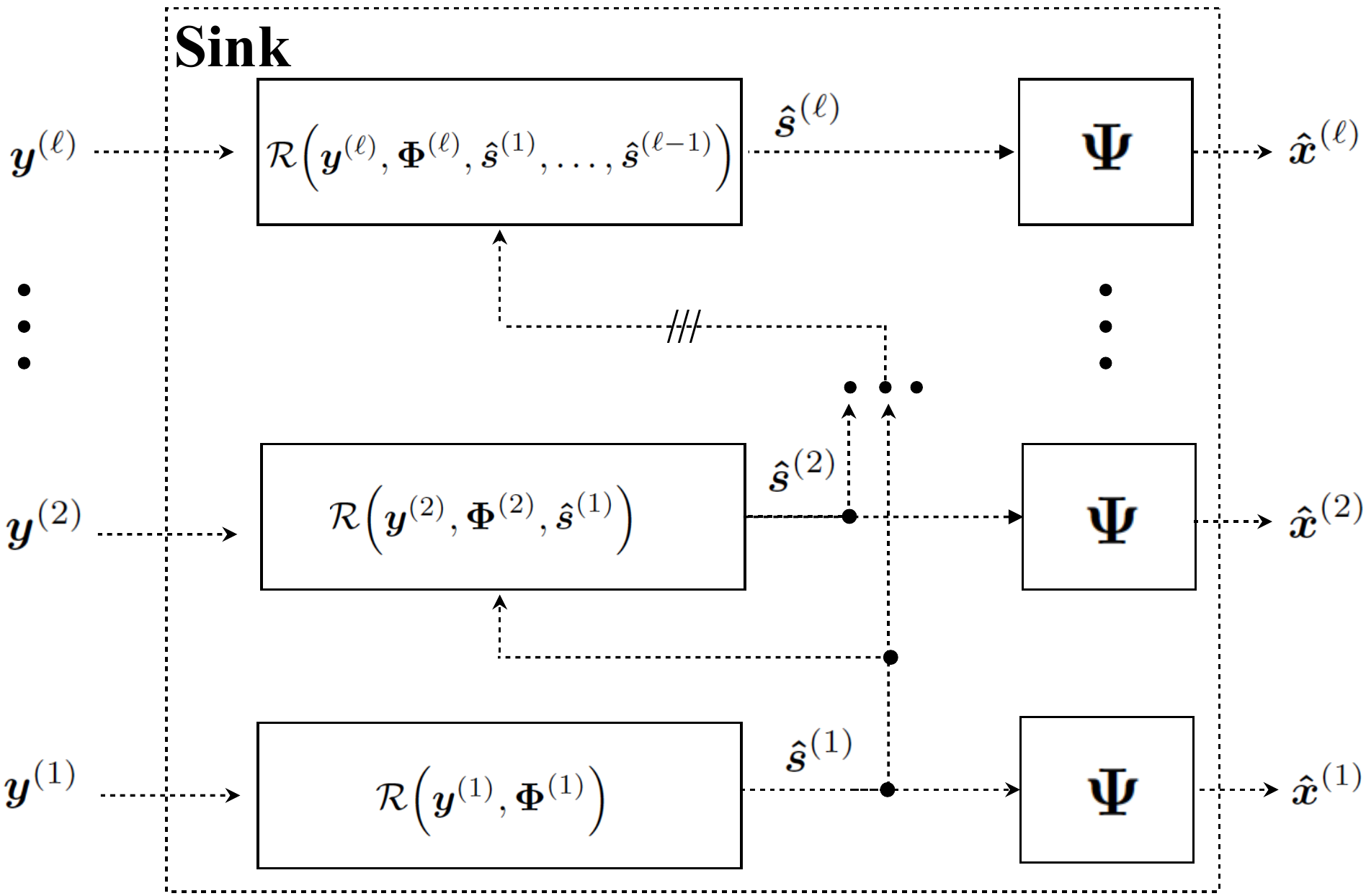}  
\caption{
  Diagram of the proposed data recovery scheme. The vectors of readings $\bm{x}^{(l)}$ of each data type are
  reconstructed sequentially, $l = 1, \ldots, \ell$. The reconstruction of $\bm{x}^{(l)}$ uses the
  respective measurements $\bm{y}^{(l)}$ and matrix $\bm{\Phi}^{(l)}$, as well as the sparse
  representations of the previously reconstructed modalities, $\bm{s}^{(1)}$, \ldots,
  $\bm{s}^{(l-1)}$. 
}
\label{fig:ourDecoder}
\end{figure}

\section{CS for Heterogeneous Networked Data}
\label{sec:PropDataRec}

State-of-the-art compressive data gathering and recovery  solutions~\cite{haupt2008compressed,quer2012sensing,duarte2005distributed,luo2010efficient}
leverage the spatiotemporal correlation among \textit{homogeneous} sensor readings, collected by a
given sensing modality.  However, current WSN and IoT setups  involve diverse sensing devices
gathering \textit{heterogeneous} data; for instance, different air pollution measurements (CO,
$\text{NO}_2$, $\text{O}_3$, $\text{SO}_2$) are collected in an environmental  monitoring setup.
\textcolor{black}{We propose a design that jointly reconstructs heterogeneous correlated data from
  compressive measurements, by leveraging both intra-
and inter-source data  dependencies.} 


\textcolor{black}{
Consider a network comprising $n$ wireless devices, each of which equipped with~$\ell$
sensors that monitor diverse, but statistically dependent, data types; for example, if the sensors
measure the concentration of CO, $\text{NO}_2$, $\text{O}_3$, and $\text{SO}_2$, then $\ell=4$.
Let~$x_i^{(l)}$ denote the reading at sensor $i \in \{1,\ldots,n\}$ of data type $l \in \{1, \ldots,
\ell\}$, and let~$\bm{x}^{(l)}=\big(x^{(l)}_1, \ldots, x^{(l)}_i, \ldots, x^{(l)}_n\big)$
be the vector collecting all the readings of data type~$l$. We
assume that~$\bm{x}^{(l)}$ is sparse or compressible in a given
orthonormal basis~$\bm{\Psi}$; that is, $\bm{x}^{(l)}=\bm{\Psi}\bm{s}^{(l)}$,
where~$\bm{s}^{(l)}=\big(s^{(l)}_1,\ldots, s^{(l)}_i,\ldots, s^{(l)}_n\big)$ is the compressible
representation of~$\bm{x}^{(l)}$.} In our experiments in
Section~\ref{sec:BasisSelection}, $\bm{\Psi}$ will be the discrete cosine transform (DCT) as,
among several other common transforms, this is the one that yields the sparsest representation of
pollution data. 

\textcolor{black}{
The data gathering schemes~\cite{haupt2008compressed,luo2010efficient,quer2012sensing} that were
reviewed in Section~\ref{sec:SoTA_scheme} can be readily extended to address the collection of
heterogeneous data. Fig.~\ref{fig:ourEncoder} shows how we modify the multi-hop scheme
of~\cite{luo2010efficient}. Specifically, we assume that the communication network is a line graph,
starting at node $1$ and ending at node $n$. Node $n$, in turn, is connected to the sink node. The
measurements of data type $l$ are collected and transmitted as was described in
Section~\ref{sec:SoTA_scheme}: node $1$ measures $x_1^{(l)}$, and transmits
$\phi_{1,1}^{(l)}x_1^{(l)}$ to node $2$, where $\phi_{1,1}^{(l)}$ is randomly generated; node $2$, in turn,
measures $x_2^{(l)}$, generates $\phi_{1,2}^{(l)}$, computes $\phi_{1,2}^{(l)}x_2^{(l)}$, and
transmits the sum $\phi_{1,2}^{(l)}x_2^{(l)} + \phi_{1,1}^{(l)}x_1^{(l)}$ to node $3$; and so on. The
process is repeated $m^{(l)}$ times, each time for different realizations of $\phi_{j,i}^{(l)}$. The
sink then obtains the vector of measurements for source $l$:
} 
\begin{equation}
\label{eq:DataGathering}
\bm{y}^{(l)}=\left[\bm{\phi}_{1}^{(l)}\dots\bm{\phi}_{i}^{(l)}\dots\bm{\phi}_{n}^{(l)}\right]\cdot\bm{x}^{(l)}=\bm{\Phi}^{(l)}\bm{x}^{(l)}\,,
\end{equation}  
which has length $m^{(l)}$. 
\textcolor{black}{
Whenever the communication medium and the receiver of the sink have imperfections,
\eqref{eq:DataGathering} can be modified to
$\bm{y}^{(l)}=\bm\Phi^{(l)}\bm{x}^{(l)}+\bm{z}^{(l)}$, where
$\bm{z}^{(l)}\in\mathbb{R}^{m^{(l)}}$ is additive white Gaussian noise (AWGN)~\cite{haupt2008compressed}.
}
The collection and transmission of measurements of the other data types
is performed in the exact same way, either sequentially or concurrently.

\textcolor{black}{
Unlike the studies in~\cite{haupt2008compressed,luo2010efficient,quer2012sensing},
the sensing matrix we consider here
is~$\bm\Phi^{(l)}=\bm{\Theta}^{(l)}\bm{\Psi}^{T}$,
where~$\bm{\Theta}^{(l)}$ is a sparse Rademacher matrix~\cite{baron2010bayesian},
and~$\bm{\Psi}^T$ is the transpose of~$\bm{\Psi}$. Each measurement vector
in~\eqref{eq:DataGathering} can then be written as
$
\bm{y}^{(l)}
=
\bm{\Phi}^{(l)}\bm{x}^{(l)}
=
\bm{\Theta}^{(l)}
\bm{\Psi}^T
\bm{\Psi}\bm{s}^{(l)}
=
\bm{\Theta}^{(l)}\bm{s}^{(l)}
$, 
where~$\bm{\Psi}^T\bm{\Psi}=\bm{I}$ because $\bm{\Psi}$ is orthonormal. 
Bearing a similarity with low-density parity-check matrices\cite{gallager1962low}, sparse Rademacher matrices have 
only few non-zero entries, which are either $-1$ or $1$, with equal probability. As shown in~\cite{baron2010bayesian}, they can lead to 
accurate and fast belief-propagation-based CS decoding, as opposed to dense Gaussian 
matrices \cite{Donoho06compressed,cande2008introduction}. Similarly to the work in~\cite{baron2010bayesian}, 
the row weight $\lambda$ and the column weight $\rho$ of $\bm{\Theta}^{(l)}$ are kept very 
low---with respect to the dimension of the row and the column, respectively---and are assumed to be  
constant. Note that our selection for $\bm{\Phi}^{(l)}$ requires all nodes to know the matrices
$\bm{\Theta}^{(l)}$ and $\bm{\Psi}^{(l)}$, which can be accomplished by having all the nodes share a
seed for generating the random entries of $\bm{\Theta}^{(l)}$; and, if required, the matrix $\bm{\Psi}^{(l)}$
can easily be pre-stored.} 

\textcolor{black}{
After receiving the measurements $\bm{y}^{(l)}$ for all data types~$l=1,\dots,\ell$, the sink
then proceeds to the data recovery stage, which is the focus of our paper, and is shown schematically
in Fig.~\ref{fig:ourDecoder}.
} 
Our method operates in
stages, with the sink reconstructing the vectors~$\bm{s}^{(l)}$ sequentially, i.e., first
$\bm{s}^{(1)}$, then $\bm{s}^{(2)}$, until $\bm{s}^{(\ell)}$. When reconstructing $\bm{s}^{(l)}$, the
sink uses the measurements that were relayed, $\bm{y}^{(l)}$, the matrix $\bm{\Phi}^{(l)}$, as well as
the  previously reconstructed vectors $\bm{\hat{s}}^{(1)},\dots,\bm{\hat{s}}^{(\l-1)}$, which play
the role of multiple side information. 

\textcolor{black}{Standard CS recovery
algorithms~\cite{tropp2007signal,donoho2009message,baron2010bayesian}, as proposed
by~\cite{haupt2008compressed,luo2010efficient,quer2012sensing}, would require recovering each
sparse vector $\bm{s}^{(l)}$ independently from the other vectors, based only on the
measurement vector $\bm{y}^{(l)}$. This fails to leverage \textit{inter}-source
correlations. We will refer to this approach as the \textit{baseline solution}.
} 
Alternatively, one can apply DCS \cite{duarte2005distributed} to recover the ensemble of sparse vectors~$\{\bm{s}^{(l)}\}_{l=1}^{\ell}$  using the ensemble of measurements vectors~$\{\bm{y}^{(l)}\}_{l=1}^{\ell}$
and the matrices $\{\bm \Phi^{(l)}\}_{l=1}^{\ell}$. 
However, as shown in our experimental results, DCS does not efficiently capture the underlying
dependencies among heterogeneous data, such as various air pollutants, which have different
statistical properties. 

The method we propose, in contrast, leverages diverse correlated signals through copula
functions~\cite{sklar1959fonctions,nelsen2006introduction}. Copula functions, explained in detail in
Section~\ref{SubSec:IntroCop}, are elements of a statistical
framework to effectively capture dependencies between random variables. As will be explained in
Section~\ref{sec:prop_recovery}, we use copula functions to integrate knowledge from other
data types in the reconstruction of a given data type or, in other words, as a way to integrate multiple
side information. 
\textcolor{black}{
Our experiments in Section \ref{sec:Experiments} show that it is exactly because it uses multiple side
information signals at the recovery stage that our scheme outperforms the state-of-the-art
methods in~\cite{duarte2005distributed,luo2010efficient,mota2014compressed,mota2016ref,mota2014glob}.
}

\section{Statistical Modelling Using Copulas}
\label{sec:StatModel}

We now describe how to model statistically heterogeneous data using copula functions.
\textcolor{black}{
Let $S^{(l)}$ denote the random variable associated with the reading of source $l \in
\{1,\ldots,\ell\}$, and let $s^{(l)}$ be one of its realizations. In general, the data sources we
consider are not independent, meaning their joint probability
density function (pdf) $f_{S^{(1)}, S^{(2)}, \ldots, S^{(\ell)}}
\big(s^{(1)},s^{(2)},\ldots,s^{(\ell)}\big)$ does not factor into the product of its marginals. 
We will represent this joint pdf as $f_{\bm{S}}(\bm{s})$, where $\bm{S} := (S^{(1)}, \ldots,
S^{(\ell)})$ is a random vector and $\bm{s} := (s^{(1)}, \ldots, s^{(\ell)})$ its
realization.\footnote{Notice
the difference in notation with respect to $\bm{s}^{(l)} := (s_1^{(l)}, \ldots, s_n^{(l)})$, which
collects the samples observed from data type $l$.} We 
assume that each sensor $i \in \{1, \ldots, n\}$ observes $\bm{S}_i$, an independent realization of
$\bm{S}$. In other words, $\bm{S}_i$ is an i.i.d.\ copy of $\bm{S}$. 
} 
This implies  
\begin{align}
f_{\bm{S}_1,\ldots,\bm{S}_n} \big(\bm{s}_1,\ldots,\bm{s}_n\big)
&=
\prod_{i=1}^n  f_{\bm{S}_i}(\bm{s}_i)
\notag
\\
&=
\prod_{i=1}^n f_{S_i^{(1)},\ldots,S_i^{(\ell)}}
\Big(s_i^{(1)}\ldots,s_i^{(\ell)}\Big)\,.
\label{Eq:DecompS}
\end{align} 
We will see next how a copula function enables working with the marginals of the joint pdfs
$f_{S_i^{(1)},\ldots,S_i^{(\ell)}} \big(s_i^{(1)},\ldots,s_i^{(\ell)}\big)$ in~\eqref{Eq:DecompS},
even though, as we saw before, these pdfs do not factor into the product of their marginals.


\subsection{Introduction to Copulas} 
\label{SubSec:IntroCop}

\textcolor{black}{
Suppose the random vector $\bm{S} = \big(S^{(1)}, \ldots, S^{(\ell)}\big)$ is supported on a
continuous set $\bm{\mathcal{S}} \subseteq \mathbb{R}^{\ell}$ and has joint cumulative distribution
function (cdf)}
\begin{equation*}
F_{S^{(1)},\ldots,S^{(\ell)}}(s^{(1)},\ldots,s^{(\ell)})
=
\text{Pr}\big[S^{(1)}\leq s^{(1)},\ldots,S^{(\ell)}\leq s^{(\ell)}\big]\,.
\end{equation*}
We will denote the marginal cdfs by $F_{S^{(l)}}(s^{(l)})=\text{Pr}\left[S^{(l)}\leq s^{(l)}\right]$.
\textcolor{black}{
The probability integral transform~\cite{genest2001multivariate} states that, independently of the
distribution of $S^{(l)}$, the random variable $U^{(l)} :=
F_{S^{(l)}}\big(S^{(l)}\big)$ always has uniform distribution over $[0, 1]$. 
}

\textcolor{black}{ 
The copula function of the random vector $\bm{S} = (S^{(1)}, \ldots, S^{(\ell)})$ is defined on the
unit hypercube $[0, 1]^\ell$ as the joint cdf of $\bm{U}:=(U^{(1)}, \ldots, U^{(\ell)})$, that is,
\begin{align}
C(u^{(1)},\cdots,u^{(\ell)}) 
= \text{Pr}\Big[U^{(1)}\leq u^{(1)},\ldots,U^{(\ell)}\leq u^{(\ell)} \Big],
\end{align}
where $u^{(l)} = F_{S^{(l)}}(s^{(l)})$. Namely, a copula is a multivariate cdf whose marginals have
uniform distribution. The following result was seminal in the development of the theory of copula
functions. 
}
\begin{theorem}[Sklar's theorem~\cite{sklar1959fonctions}]
  \label{Thm:Sklar}
For any~$\ell$-dimensional joint cdf~$F_{S^{(1)},\ldots,S^{(\ell)}}(s^{(1)},\ldots,s^{(\ell)})$ whose
marginals are continuous, there exists a unique $\ell$-dimensional copula function~$C:[0,1]^\ell \rightarrow
[0,1]$ such that
\begin{equation}
F_{S^{(1)},\ldots,S^{(\ell)}}(s^{(1)},\ldots,s^{(\ell)}) = C(u^{(1)},\ldots,u^{(\ell)}).
\label{eq:copulaCumulFuncDef}
\end{equation}
\end{theorem}
The implications of Theorem~\ref{Thm:Sklar} are best seen after taking the $\ell$-th cross partial derivative
of~\eqref{eq:copulaCumulFuncDef}:
\begin{align}
&f_{S^{(1)},\ldots,S^{(\ell)}} \big(s^{(1)},\ldots,s^{(\ell)}\big)
=
\frac{\partial F_{S^{(1)},\ldots,S^{(\ell)}}(s^{(1)},\ldots,s^{(\ell)}) }{\partial u^{(1)} \cdots
  \partial u^{(\ell)}}
\nonumber 
\\
&= c\big(u^{(1)},\ldots,u^{(\ell)}\big) \times \prod_{l=1}^{\ell} f_{S^{(l)}}\big(s^{(l)}\big),
\label{eq:copulaPDFDef}
\end{align} 
where~$c(u^{(1)},u^{(2)},\dots,u^{(\ell)})=\frac{\partial^nC(u^{(1)},u^{(2)},\dots,u^{(\ell)})}{\partial
  u^{(1)},\partial u^{(2)},\dots,\partial u^{(\ell)}}$ denotes the copula density, and~$f_{S^{(l)}}(s^{(l)})$ is the pdf of $S^{(l)}$. 
\textcolor{black}{
Expression~\eqref{eq:copulaPDFDef} tells us that the joint pdf of \textit{dependent} random variables
can be written as the product of the marginal pdfs, as if the variables were \textit{independent},
times the copula density, which acts as a correction term. In other words, the copula density
alone captures all the dependencies of the random variables.
This means that finding a good model for the joint pdf boils down to finding not only accurate models
for the marginal pdfs, but also an appropriate
copula function to effectively capture the dependencies in the data.
} 

\subsection{Copula Families} 
\label{sec:CopulaFamilies}

There exist several bivariate and multivariate copula
families~\cite{joe1997multivariate,genest1986joy,nelsen2006introduction}, typically categorized into
\textit{implicit} and \textit{explicit}. Implicit copulas have densities with no simple closed-form expression, but  are derived from well known distributions. An example is the Elliptical
copulas, which are associated to elliptical distributions (for example, the multivariate normal
distribution), and have the advantage of providing
symmetric densities. This makes them appropriate for high-dimensional distributions.
Table \ref{tab:Elliptical} shows the expressions for the two mostly used Elliptical copulas: the
Gaussian, and the Student's $t$-copula \cite{sklar1959fonctions}. 
The expression for the Gaussian copula uses a standard multivariate normal distribution parameterized
by the correlation matrix~$\bm{R}_G$. In turn, the expression for the Student's $t$-copula
uses a standard multivariate $t$-distribution, parameterized by the correlation
matrix~$\bm{R}_t$ and by the degrees of freedom $\nu$. 
\textcolor{black}{The diagonal entries of the correlation matrices~$\bm{R}_{(\cdot)}$ are $1$,
and the non-diagonal are equal to the estimated Spearman's $\rho$ values.}

%
%
%
%
%

\begin{table*}[t]
\caption{Elliptical Copula Functions}
\label{tab:Elliptical}
\centering 
\footnotesize
\begin{tabular}{|c|c|c|c|} 
\hline
Name & $C_e\Big(u^{(1)},\dots,u^{(\ell)}\Big)$ & Parameters    & Functions \\ [0.15ex]\hline\hline
\multirow{ 2}{*}{Gaussian}       & \multirow{ 2}{*}{$\Phi_{\bm{R}_G}\big(\Phi_g^{-1}(u^{(1)}),...,\Phi_g^{-1}(u^{(\ell)})\big)$}
 & \multirow{ 2}{*}{$\bm{R}_G$: correlation matrix}   & $\Phi_{\bm{R}_G}:$ standard multivariate
normal distribution\\
& &   & $\Phi_g:$ standard univariate normal distribution\\\hline\hline
\multirow{ 2}{*}{Student}  & \multirow{ 2}{*}{$T_{\bm{R}_t,\nu} \big(t_{\nu}^{-1}(u^{(1)}),...,t_{\nu}^{-1}(u^{(\ell)})\big)$}
 & $\bm{R}_t:$ correlation matrix     & $T_{\bm{R}_t,\nu}:$ standard multivariate $t$-distribution\\
 & & $\nu:$ degrees of freedom & $T_{\nu}:$ univariate $t$-distribution\\ [0.1ex]\hline
\end{tabular}
\end{table*}

Explicit copulas have  densities with simple closed-form expressions but, being typically
parameterized by few parameters,  lack some modeling flexibility. The most popular explicit
copulas are the Archimedean, which are parameterized by a single parameter $\xi \in \Xi \subseteq
\mathbb{R}$. Specifically, an Archimedean copula is defined as \cite{nelsen2006introduction}:
\begin{equation}
  \label{Eq:ArchimedeanCopula}
C_{a}(u^{(1)},\dots,u^{(\ell)};\xi)=q^{-1}\Big(q(u^{(1)};\xi)+\dots+q(u^{(\ell)};\xi);\xi\Big),
\end{equation}
where $q:[0,1]\times \Xi \rightarrow [0,\infty )$ is a continuous,
strictly decreasing, convex function such that $q(1;\xi)=0$. The
function $q(u)$ is called generator and its pseudo-inverse, defined
by
\begin{equation}
q^{-1}(u;\xi)
=
\left\{
  \begin{array}{ll}
    q(u;\xi) & \text{if $0\leq u \leq q(0;\xi)$} 
    \vspace{0.1cm}
    \\
    0        & \text{if $q(0;\xi)\leq u\leq \infty$}\,,
  \end{array}
\right.\,
\end{equation}
has to be strictly-monotonic of order~$\ell$~\cite{mcneil2009multivariate}. Table
\ref{tab:Archimedean} shows the distributions of the most popular Archimedean copulas: the
Clayton, the Frank, and the Gumbel copulas~\cite{mcneil2015quantitative}. 

For both families, the estimation of the copula parameters, e.g., the correlation matrix, is
performed using training data. This will be described in detail in Section~\ref{sec:Experiments}.
    
\begin{table*}[t]
\caption{Archimedean Copula Functions}
\label{tab:Archimedean}
\centering 
\footnotesize
\begin{tabular}{|c|c|c|c|} 
\hline
Name & $C_a\Big(u^{(1)},\dots,u^{(\ell)}\Big)$   & Parameter Range $\Xi$    & Generator
$q(u)$ \\ [0.15ex]\hline\hline
\multirow{ 2}{*}{Clayton}       & \multirow{ 2}{*}{$\left(\sum_{l=1}^{\ell}(u^{(i)})^{-\xi}-\ell+1\right)^{-\sfrac{1}{\xi}}$}
 & \multirow{ 2}{*}{$\xi\in(0,\infty)$}   & \multirow{ 2}{*}{$\xi^{-1}\left(u^{-\xi}-1\right)$}\\
& &   & \\\hline\hline
\multirow{3}{*}{Frank}  & \multirow{3}{*}{$-\frac{1}{\xi}\log\left(1+\frac{\prod_{l=1}^{\ell}\left(e^{-\xi
u^{(l)}}-1\right)}{\left(e^{-\xi}-1\right)^{\ell-1}}\right)$}
 & \multirow{3}{*}{$\xi \in (-\infty,\infty)$}  & \multirow{3}{*}{$-\log\left(\frac{e^{-\xi
u}-1}{e^{-\xi}-1}\right)$}\\
 & &  & \\ & & & \\ [0.1ex]\hline\hline
\multirow{3}{*}{Gumbel}  & \multirow{3}{*}{$\exp\left[\left(-\sum_{l=1}^{\ell}(-\log
u^{(l)})^{\xi}\right)^{\sfrac{1}{\xi}}\right]$} & \multirow{3}{*}{$\xi
\in [1,\infty)$} & \multirow{3}{*}{$(-\log u)^{-\xi}$}\\
& & & \\ & & & \\[0.1ex]\hline
\end{tabular}
\end{table*}

\subsection{Marginal Statistics}
\label{sec:MarginalStatistics}

As shown in~\eqref{eq:copulaPDFDef}, a consequence of Sklar's theorem (Theorem~\ref{Thm:Sklar}) is
that copula functions enable us to work with the marginal pdfs of a random vector even when its components are not
independent. We will consider the following pdfs when we model the distribution of each component. 
\begin{enumerate}
  \item 
    Laplace distribution 
    \begin{equation}
      \label{Eq:LaplaceDistribution}
      f_{S^{(l)}}\big(s^{(l)};b^{(l)}\big)
      =
      \frac{1}{2b^{(l)}}\exp \bigg[-\frac{\big|s^{(l)}-\mu^{(l)}\big|}{b^{(l)}}\bigg],
    \end{equation}
    where $b^{(l)}$ is the scaling parameter and $\mu^{(l)}$ is the mean value for the~$l$-th data
    type, with~$l\in\{1,2,\dots,\ell\}$.

  \item 
    Cauchy (or Lorentz) distribution 
    \begin{equation}
      \label{Eq:CauchyDistribution}
      f_{S^{(l)}}\big(s^{(l)};\alpha^{(l)},\beta^{(l)}\big)
      =
      \frac{1}{\pi\beta^{(l)}}\Bigg[1+\bigg(\frac{s^{(l)}-\alpha^{(l)}}{\beta^{(l)}}\bigg)^2\Bigg]^{-1}\!\!,
    \end{equation}
    where $\beta^{(l)}$ is a scale parameter specifying the half-width at half-maximum, and
    $\alpha^{(l)}$ is the location parameter. 
  \item 
    \textcolor{black}{Non-parametric distribution via kernel density estimation (KDE)~\cite{roussas2003} 
      \begin{equation}
        f_{S^{(l)}}\big(s^{(l)};h^{(l)}\big)
        =
        \frac{1}{n\cdot h^{(l)}}\sum_{i=1}^{n}\mathcal{K}\left(\frac{s^{(l)}-s^{(l)}_{i}}{h^{(l)}}\right),
        \label{eq:kde}
      \end{equation} 
      where $n$ is the number of samples from data type $l\in\{1,2,\dots,\ell\}$. We use the Gaussian
      kernel $\mathcal{K}(v)=\frac{1}{\sqrt{2\pi}}\exp{\left(-\frac{1}{2}v^2\right)}$ because of its
      simplicity and good fitting accuracy. We also select different smoothing parameters
      $h^{(l)}$ for different data types, $l \in \{1, \ldots, \ell\}$.
    } 
\end{enumerate}

\section{Copula-based Belief Propagation}
\label{sec:prop_recovery}

We now describe our reconstruction algorithm, executed at the sink node. As mentioned, the sparse
vectors $\bm{s}^{(l)}$ are reconstructed sequentially: first,
$\bm{s}^{(1)}$, then $\bm{s}^{(2)}$, and so on. The reconstruction of each $\bm{s}^{(l)}$ thus uses
not only the respective measurements $\bm{y}^{(l)}$, but also the previously reconstructed
data types $\hat{\bm{s}}^{(1)},\dots,\hat{\bm{s}}^{(l-1)}$ as side information.

\textcolor{black}{
We adopt the framework of Bayesian CS~\cite{baron2010bayesian,ji2008bayesian}, as it naturally
handles our joint statistical characterization of the correlated modalities. We start by
computing the posterior distribution of the random vector $\bm{S}^{(l)}$, representing the sparse
vectors of coefficients of data type $l$, given the respective measurements $\bm{Y}^{(l)}$ and the 
first $l-1$ data types:
}
\begin{align}
    f&_{  
      \bm{S}^{(l)}
      \vert
      \bm{Y}^{(l)}
      \bm{S}^{(1)}
      \cdots
      \bm{S}^{(l-1)}
    }
  \label{Eq:Posterior1}
  \\
  &
  \propto 
    f_{
     \bm{Y}^{(l)}\vert
     \bm{S}^{(1)}
     \cdots
     \bm{S}^{(l)}
    }
   \times 
    f_{
      \bm{S}^{(l)}
      \vert
      \bm{S}^{(1)}
      \cdots
      \bm{S}^{(l-1)}
    }
  \label{Eq:Posterior2}
  \\ 
  &=
    f_{ 
      \bm{Y}^{(l)}
      \vert
      \bm{S}^{(l)}
    }
  \times
    f_{
      \bm{S}^{(l)}
      \vert
      \bm{S}^{(1)}
      \cdots
      \bm{S}^{(l-1)}}
  \label{Eq:Posterior3}
  \\
  &= 
    \prod_{j=1}^{m^{(l)}}
    f_{
      Y_j^{(l)}
      \vert
      \bm{S}^{(l)}
    }
  \times
    \prod_{i=1}^{n} 
    f_{
      S^{(l)}_i
      \vert
      {S}^{(1)}_i
      \cdots
      {S}_i^{(l-1)}
    }
    \,, 
  \label{eq:app}
\end{align}
\textcolor{black}{
where we excluded the arguments of the pdfs for notational simplicity. From~\eqref{Eq:Posterior1}
to~\eqref{Eq:Posterior2}, we just applied Bayes's theorem and omitted constant terms. From~\eqref{Eq:Posterior2}
to~\eqref{Eq:Posterior3}, we used the assumption that measurements from data type $l$ given realizations of
all the previous data types $j \leq l$ depend only on the value of $\bm{S}^{(l)} = \bm{s}^{(l)}$; in other
words, the process $\bm{Y}^{(l)}\vert \bm{S}^{(1)}\cdots \bm{S}^{(l)} = \bm{Y}^{(l)}\vert
\bm{S}^{(l)}$ is Markovian. Finally, from~\eqref{Eq:Posterior3} to~\eqref{eq:app}, we used the
assumption that measurement noise at different sensors is independent, and also that each sensor
observes independent realizations of the random vector $\bm{S} = (S^{(1)}, \ldots, S^{(\ell)})$ (cf.\
Section~\ref{sec:StatModel}).
} 
Obtaining an estimate of~$\bm{s}^{(l)}$ by minimizing the mean-squared-error or via maximum a
posteriori (MAP) is challenging due to the complexity of the posterior distribution in
\eqref{eq:app}. Therefore, as in \cite{baron2010bayesian}, we use the belief propagation
algorithm~\cite{mackay2003information}. 

\begin{figure}[t]
\centering
\includegraphics[scale=0.9]{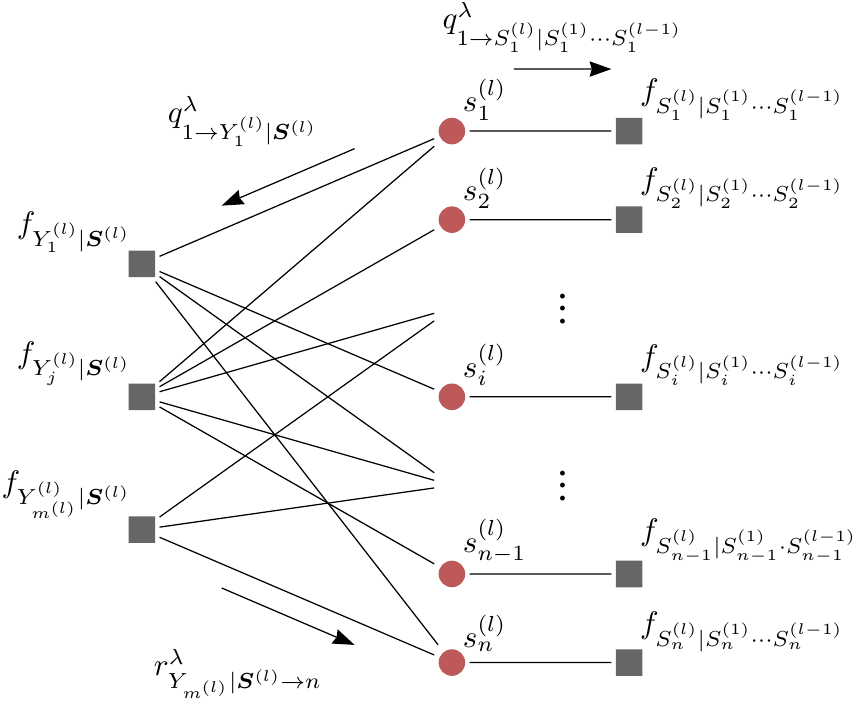}               
\caption{
  Factor graph corresponding to the posterior distribution~\eqref{eq:app}. The variable nodes are
  represented with circles and the factor nodes with squares. A message from variable node $s_i^{(l)}$ to factor
  node $f_Z$ at iteration $\lambda$ is denoted with $q^\lambda_{i\rightarrow Z}$, and a message in the 
  inverse direction is denoted with $r^\lambda_{Z \rightarrow i}$.  
}
\label{fig:tanner_dec} 
\end{figure}

Our approach modifies the algorithm in \cite{baron2010bayesian}
to take into account 
the previously reconstructed signals~$\bm{s}^{(1)},\ldots,\bm{s}^{(l-1)}$ 
in the reconstruction of $\bm{s}^{(l)}$.
Fig.~\ref{fig:tanner_dec} represents the factor graph associated with~\eqref{eq:app}. A factor graph
represents the factorization of an expression by using two types of nodes: \textit{variable nodes}
and \textit{factor nodes}. Variable nodes are associated to the variables of the expression, in this
case, the components of the vector $\bm{S}^{(l)} = \big(s_1^{(l)}$, \ldots, $s_n^{(l)}\big)$ and, in
Fig.~\ref{fig:tanner_dec}, are represented with circles. The factor nodes are associated to the
intermediate factors in the expression, in this case, the terms in~\eqref{eq:app} and, in
Fig.~\ref{fig:tanner_dec}, are represented with squares. Specifically, the leftmost squares in the
figure represent the terms in the product  $\prod_{j=1}^{m^{(l)}} f_{ Y_j^{(l)} \vert
\bm{S}^{(l)} }$, and the rightmost squares represent terms in the product $\prod_{i=1}^{n}
f_{ S^{(l)}_i \vert {S}^{(1)}_i \cdots {S}_i^{(l-1)} }$. 

Notice that when there is no measurement noise in the acquisition of the measurements $\bm{y}^{(l)}$,
each factor 
$f_{Y_j^{(l)}\vert\bm{S}^{(l)}}$ becomes
\begin{equation}
  \label{Eq:Dirac}
    f_{Y_j^{(l)}\vert\bm{S}^{(l)}}
    \Big(y_j^{(l)}\vert\bm{s}^{(l)}\Big)
  =
    \delta
    \Big(
      y_j^{(l)} -\sum_{i=1}^N\Theta_{j,i}^{(l)}\,s_i^{(l)}
    \Big)\,,
\end{equation}
where $\delta(\cdot)$ is the Dirac delta function. When the measurement noise is, for example, AWGN,
then $\delta(\cdot)$ in~\eqref{Eq:Dirac} is replaced by the density of the normal distribution.  
Therefore, in Fig.~\ref{fig:tanner_dec}, the edges from the variable nodes to the leftmost factor
nodes represent the connections defined by measurement equation
$\bm{y}^{(l)}=\bm{\Theta}^{(l)}\bm{s}^{(l)}$ (cf.\
Section~\ref{sec:PropDataRec}): there is an edge between factor $f_{Y^{(l)}_j\vert \bm{S}^{(l)}}$ and
variable $s^{(l)}_i$ whenever $\Theta_{ji}^{(l)} \neq 0$. Recall also that the nonzero entries of
$\bm{\Theta}^{(l)}$ are
$\pm 1$.

Regarding the connections with the rightmost factor nodes in Fig.~\ref{fig:tanner_dec}, notice
that~\eqref{eq:copulaPDFDef} implies that each term $f_{ S^{(l)}_i \vert {S}^{(1)}_i \cdots
  {S}_i^{(l-1)} }$ can be expressed as the marginal pdf $f_{ S^{(l)}_i}$ times a correction term that
captures information from the previously reconstructed data types. Indeed, assuming we have access to
estimates $\widehat{\bm{s}}^{(k)}$ of $\bm{s}^{(k)}$, for $k<l$, there holds
\begin{align}
  & 
    f_{S^{(l)}_i\vert{S}^{(1)}_i,\dots,{S}_i^{(l-1)}}
    \Big(s^{(l)}_i\,\Big\vert\, \widehat{s}^{(1)}_i,\ldots,\widehat{s}^{(l-1)}_i\Big)
  \label{Eq:CopDensity1} 
  \\
  &=
    \frac{
      f_{{S}^{(1)}_i,\ldots,{S}_i^{(l)}}
      \Big(s^{(l)}_i,  \widehat{s}^{(1)}_i,\ldots,\widehat{s}^{(l-1)}_i\Big)
    }
    {
      f_{S_i^{(1)},\ldots, S_i^{(l-1)}}
      \Big(\widehat{s}^{(1)}_i,\ldots,\widehat{s}^{(l-1)}_i\Big)
    }
  \label{Eq:CopDensity2}
  \\
  &=
    \frac{
      c\big(\hat{u}^{(1)}_i,\ldots,\hat{u}^{(l-1)}_i,u_{i}^{(l)}\big)
    }
    { 
      c\big(\hat{u}^{(1)}_i,\ldots,\hat{u}^{(l-1)}_i\big)
    }
    \cdot
    f_{S^{(l)}_i}\big(s_i^{(l)}\big)\,,
  \label{eq:copula_init_message}
\end{align}
where $\hat{u}_i^{(k)} = F_{S_i^{(k)}}\big(\hat{s}_i^{(k)}\big)$ for $k = 1, \ldots, l - 1$. 
From~\eqref{Eq:CopDensity1} to~\eqref{Eq:CopDensity2} we used the definition of conditional density,
and from~\eqref{Eq:CopDensity2} to~\eqref{eq:copula_init_message} we simply
used~\eqref{eq:copulaPDFDef}. Expression~\eqref{eq:copula_init_message} depends only on $s_i^{(l)}$
and thus explains the edges from
the variables nodes to the rightmost factor nodes in Fig.~\ref{fig:tanner_dec}.

Belief propagation is an iterative algorithm in which each variable node $s^{(l)}_i$ sends a message
to its neighbors $\mathcal{M}_i$ (which are only factor nodes), and each factor node $f_Z$ sends a
message to its neighbors $\mathcal{N}_Z$ (which are only variable nodes). Here, $Z$ represents either
$Y_j^{(l)}\vert\bm{S}^{(l)}$, for $j = 1, \ldots, m^{(l)}$, or $S_i^{(l)}\vert S_i^{(1)} \cdots
  S_i^{(l-1)}$, for $i = 1, \ldots, n$. 
In our case, a belief propagation message is a vector
that discretizes a continuous probability distribution. For example, suppose the domain of the pdfs
is $\mathbb{R}$, but we expect the values of the variables to be concentrated around $0$. We can
partition $\mathbb{R}$ into $10$ bins around $0$, e.g., $(-\infty,-4] \cup (-4, -3] \cup \cdots \cup
(3, 4] \cup (4, +\infty)$. The message, in this case, would be a $10$-dimensional vector
whose entries are the probabilities that a random variable belongs to the respective bin. For
instance, all the messages to and from variable node $s_1^{(l)}$ are vectors of probabilities,
$\Big(\mathbb{P}\{S_1^{(l)} \in (-\infty, -4)\},  \ldots , \mathbb{P}\{S_1^{(l)} \in (4, +\infty)\}
\Big)$, which are iteratively updated and represent our belief for the (discretized) pdf of
$s_1^{(l)}$. Note, in particular, that all vectors have the same length and that all the
messages to and from a variable node $s_i^{(l)}$ depend on that variable only. 
We  represent a message from variable $s_i^{(l)}$ to factor $f_Z$ at iteration
$\lambda $ as $q_{i\rightarrow Z}^{\lambda}(s_i^{(l)})$, and a message from factor
$f_Z$ to variable $s_i^{(l)}$ as $r^\lambda_{Z \rightarrow
  i}(s_i^{(l)})$. The messages are updated as follows:\footnote{See, e.g., \cite{baron2010bayesian} for
  a more detailed account on belief propagation algorithms, including a derivation of these formulas.
Note also that, for simplicity, we omit normalizing constants.} 
\begin{align} 
    q_{i \rightarrow Z}^{\lambda}\big(s^{(l)}_i\big)
  &=
    \prod_{U \in \mathcal{M}_i \setminus \{Z\}} r_{U\rightarrow i}^{\lambda-1}\big(s^{(l)}_i\big)
  \label{eq:coefficients_message}
  \\
    r_{Z\rightarrow i}^{\lambda}\big(s^{(l)}_i\big)
  &=
    \sum_{\sim s_i^{(l)}} f_{Z}\big(Z\big) 
    \cdot\!\!
    \prod_{k \in \mathcal{N}_{Z}\setminus \{s_i^{(l)}\}} 
    q_{k\rightarrow Z}^{\lambda-1}\big(s^{(l)}_{k}\big)\,,
    \label{eq:measurements_message}
\end{align}
where $\sum_{\sim s_i^{(l)}}$ denotes the sum over all variables but $s_i^{(l)}$, and a ``product''
between messages is the pointwise product between the respective vectors.

We run the message passing algorithm \eqref{eq:coefficients_message}-\eqref{eq:measurements_message}
for $\Lambda$ iterations. To obtain the final estimate $\widehat{s}_{i}^{(l)}$ of each $s_{i}^{(l)}$,
we first compute the vector 
\begin{equation*}
  g\big(s_i^{(l)}\big)
  :=
  \prod_{U \in\mathcal{M}_{i}} 
  r_{U\rightarrow i}^{(\Lambda)}\big(s^{(l)}_i\big)\,,
\end{equation*}
and select $\widehat{s}_{i}^{(l)}$ as the mid-value of the bin corresponding to the largest entry of
$g\big(s_i^{(l)}\big)$. This gives us each component of the estimated vector of coefficients
$\widehat{\bm{s}}^{(l)}$. In turn, the estimated readings are computed as
$\widehat{\bm{x}}^{(l)}=\bm\Psi\widehat{\bm{s}}^{(l)}$.


\section{Experiments}
\label{sec:Experiments}

\textcolor{black}{We evaluate the data recovery performance of the proposed copula-based design using synthetic data (cf. Section~\ref{sec:resultsSyntheticData}) as well as actual sensor readings taken from the air pollution
database of the US Environmental Protection Agency (EPA)~\cite{epa} (cf. Section~\ref{sec:resultsRealData}). Furthermore, in Section~\ref{sec:EnergyConsumption}, we study the impact of the proposed method on the energy consumption of the wireless devices.}

\begin{figure}[t]
\centering
\subfigure[]{%
  \includegraphics[scale=0.45,trim=0.5cm .1cm 0.5cm 0.5cm]{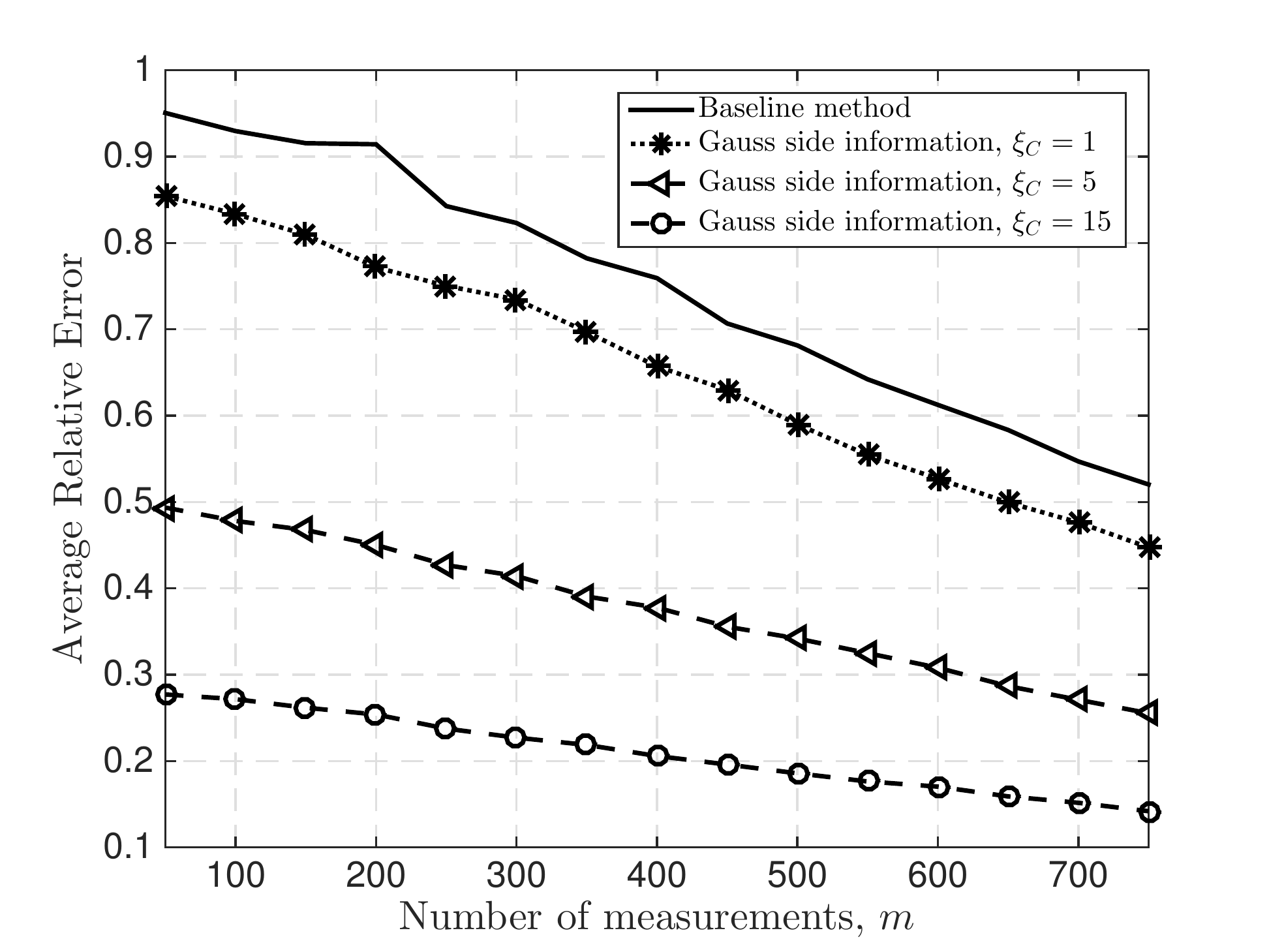}
  \label{fig:syntheticDataResults:a}}
\quad
\subfigure[]{%
  \includegraphics[scale=0.45,trim=0.5cm .1cm 0.5cm 0cm]{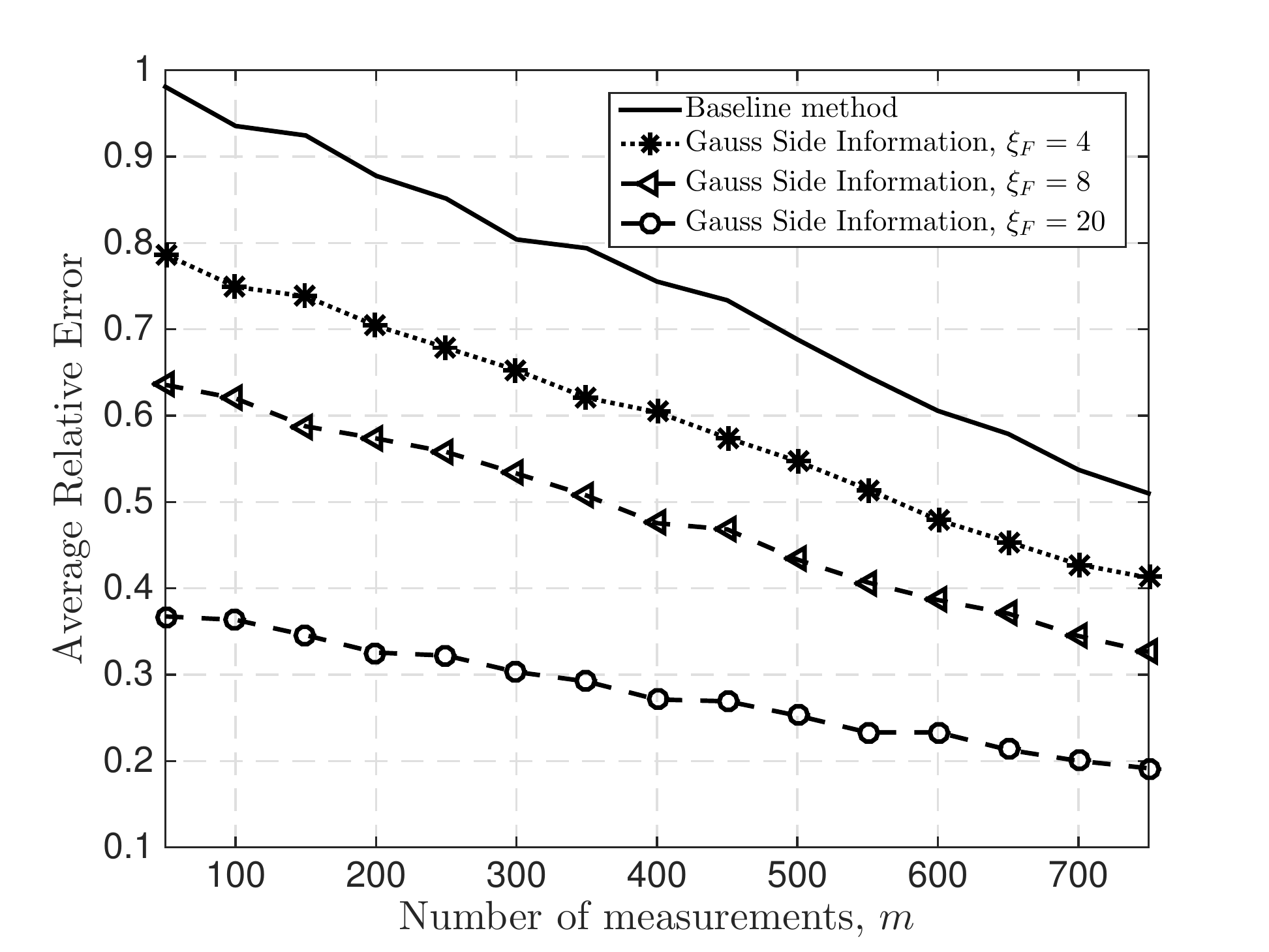}
  \label{fig:syntheticDataResults:b}} 
\caption{\textcolor{black}{Performance comparison of the proposed system against the baseline system using synthetic data. The marginal densities of the target  and side information data follow the Laplace and Gaussian distribution, respectively. The generation of the data is done using  (a) the Clayton or (b) the Frank copula function. The strength of the dependency is varied via controlling the $\xi$ parameter of the copulas.}}
\label{fig:syntheticDataResults}
\end{figure}

\subsection{\textcolor{black}{Results on Synthetic Data}}
\label{sec:resultsSyntheticData}
\textcolor{black}
{
 In order to evaluate  the proposed copula-based
method, we simulate the approach described in Section~\ref{sec:PropDataRec}. We consider the vectorized readings~$\bm{x}^{(1)}, \bm{x}^{(2)}\in\mathbb{R}^{n\times1}$ of two   statistically dependent data types collected at a given time instance by a WSN and their compressible representations~$\bm{s}^{(1)},
\bm{s}^{(2)}\in\mathbb{R}^{n\times1}$   in a basis~$\bm{\Psi}$. 
Following existing stochastic models~\cite{zordan2011modeling} for the generation of spatially-correlated WSN data, we assume that both~$\bm{x}^{(1)}$ and $\bm{x}^{(2)}$ are Gaussian. We also assume that~$\bm{x}^{(1)}$ is stationary (its variance is constant across readings), while~$\bm{x}^{(2)}$ is piece-wise stationary (its variance varies across groups of readings). Taking~$\bm{\Psi}$ as the DCT basis, it can be fairly assumed that  the coefficients in~$\bm{s}^{(1)}$ are Gaussian, whereas  the coefficients in~$\bm{s}^{(2)}$ follow the Laplace distribution\footnote{As shown in~\cite{lam2000mathematical}, the Laplace distribution emerges under the assumption that the variance  across the group of readings is exponentially
distributed.}~\cite{lam2000mathematical}.  
}
\textcolor{black}
{
To simulate this scenario, we generate~$\bm{s}^{(1)}$,~$\bm{s}^{(2)}$ as follows: We draw two coupled i.i.d. uniform random vectors~$\bm{u}^{(1)},\bm{u}^{(2)}$, with~$\bm{u}^{(l)}\in[0,1]^n$, from the bivariate Clayton or Frank copula~\cite{mcneil2015quantitative}. The length of each uniform random vector is~$n=1000$ and the copulas are parameterized by $\xi_C$ and $\xi_F$, respectively. We  consider  different values
for   $\xi_C=\{1,5,15\}$ and $\xi_F=\{4,8,20\}$,
corresponding to weak, moderate, and strong dependency,
respectively. 
We then generate the entries of $\bm{s}^{(1)}$ by applying the inverse cdf of~$\mathcal{N}(0,\sigma^2)$ with~$\sigma = 4$ to the entries of $\bm{u}^{(1)}$; similarly, the entries of $\bm{s}^{(2)}$ are generated by applying the inverse cdf of~$\mathcal{L}(0,b)$ with $b=2$ to the entries of $\bm{u}^{(2)}$.
We  obtain measurements~$\bm{y}^{(2)}=\bm{\Theta}^{(2)}\bm{s}^{(2)}$---the  column weight of~$\bm{\Theta}^{(2)}$ is set to~$\rho=20$---and we
assess the reconstruction of $\bm{s}^{(2)}$. We vary the number of measurements $m^{(2)}$ from $50$ to $750$ and, for
each $m^{(2)}$, we
perform $50$ independent trials---each with a different~$\bm{\Theta}^{(2)}$and~$\bm{s}^{(2)}$---and we report
the average relative
error~$\|\bm{s}^{(2)}-\widehat{\bm{s}}^{(2)}\|_2/\|\bm{s}^{(2)}\|_2$ as a
function of $m^{(2)}$.
}

\textcolor{black}
{
We compare the recovery performance of two methods:  the \textit{baseline method}---which recovers~$\bm{s}^{(2)}$ from~$\bm{y}^{(2)}$ via Bayesian CS with belief propagation~\cite{baron2010bayesian}---and the \textit{proposed copula-based method} that recovers~$\bm{s}^{(2)}$ using~$\bm{y}^{(2)}$~and~$\bm{s}^{(1)}$. In both methods, the length
of each message vector carrying the pdf samples in the belief propagation algorithm is set to $243$ and the number of iterations to 50. In order to have a fair comparison with CS, we account for a copula mismatch in our method. Namely, we use the bivariate
Gaussian copula  to model the dependency between the data, where the correlation matrix~$\bm{R}_G$ is fitted on the generated data using maximum likelihood estimation~\cite{bouye2000copulas}, even though the true relation between data types is generated with the Clayton or Frank copula.  
}

\textcolor{black}
{
 The experimental results, depicted
in Fig.~\ref{fig:syntheticDataResults}, show that---despite the copula mismatch---the proposed algorithm manages to leverage the dependency among the diverse data  and thus, to systematically improve the reconstruction performance  compared to   the classical method
\cite{baron2010bayesian}.  The  performance improvements  are increasing with the amount of  dependency between the  signals, reaching average relative error reductions of up to 72.90\% and 64.09\%, for the Clayton  ($\xi_{C}=15$) and the Frank copula ($\xi_{F}=20$), respectively.
}

\subsection{\textcolor{black}{Results on Real Air Pollution Data}}
\label{sec:resultsRealData}
\textcolor{black}{
The AQS (Air Quality System) database of EPA~\cite{epa} aggregates air quality
measurements taken by more than~$4000$ monitoring stations, which collect hourly or daily
measurements of the concentrations of six pollutants: ozone~$(\text{O}_3)$, particulate matter
(PM10 and PM2.5), carbon monoxide (CO), nitrogen dioxide~$(\text{NO}_2)$, sulfur
dioxide~$(\text{SO}_2)$, and lead (Pb). We consider a network architecture comprising a sink
and~$n=1000$ nodes, where each node is equipped with~$\ell=3$ sensors to measure the concentration
of CO, NO$_2$ and SO$_2$ in the air. Using the node coordinates in the EPA database, we simulate 
such networks\footnote{Each network is formed by nodes within only one
  of the following  states: CA, NV, AZ, NC, SC, VA, WV, KY, TN, MA, RI, CT,
NY, NJ, MD.} by assuming that the transmission adheres
to LoRa~\cite{lorawan}, according to which the
node distance does not exceed $2$km in urban areas and $22$km in
rural areas. From the database, we take $2\times10^5$ values for each of the three pollutants---i.e., CO, NO$_2$ and SO$_2$---collected during the year 2015. The data are equally divided into a training and an evaluation set, without overlap.
}
 

\begin{table}
\caption{Average Percentage of the Number Coefficients of the Data with an Absolute Value Below a Given Threshold~$\tau$.} 
\label{tab:TransformEvaluation}
\centering 
\tabcolsep=0.1cm
\begin{tabular}{|c|c|c|c|c|c|} 
\hline
 & $\tau$ & DCT & Haar & Daubechies-2 & Daubechies-4   \\[0.1ex]
\hline\hline
\multirow{3}{*}{$\text{SO}_2$} 
 &0.1 & \textbf{28.8} & 19.1 & 25.50 & 23.80 \\ 
  &  $0.2$ & \textbf{48.50} &  35.80 & 46.30 & 44.00\\
  &  $0.4$ & 74.10 & 67.00  & \textbf{77.50} & 77.30 \\
\hline
\multirow{3}{*}{$\text{CO}$} & 0.1 & \textbf{25.90} &  16.40 & 22.60 & 19.40\\   &  $0.2$ & \textbf{44.50} & 31.20 & 40.30 & 38.70 \\
  &  $0.4$ & \textbf{70.30} &  58.20 & 69.80 & 69.20 \\
\hline 
\end{tabular}
\end{table}

\subsubsection{\textcolor{black}{Sparsifying Basis Selection}}
\label{sec:BasisSelection}
\textcolor{black}{
We first identified a good sparsifying basis~$\bm\Psi$ for the data. Following the network
architecture described in the previous paragraph, we organized the training data into blocks
of~$n$ readings per pollutant. In order to form a block~$\bm{x}^{(l)}$, readings must have the 
same timestamp and be measured by neighboring  stations, adhering to the LoRa~\cite{lorawan} transmission 
distance criteria. We projected the data in each block onto different set of bases, including the discrete 
cosine transform (DCT), the Haar, the Daubechies-2, and the Daubechies-4 continuous wavelet transform (CWT) 
bases; for the CWT\ we experimentally found that the scale parameter~$\alpha=4$ led to the best 
compaction performance. Since the resulting representation~$\bm{s}^{(l)}$ is a compressible signal, 
we calculated the number of coefficients in~$\bm{s}^{(l)}$  whose the absolute value is below a 
given threshold~$\tau$. Table~\ref{tab:TransformEvaluation} reports the results for~$\text{SO}_2$ and CO, 
averaged over all the blocks in the training set. It shows that the DCT yielded the sparsest representations.    
}

\begin{figure*}[t]
  \centering

  \def\scaleFittings{0.282}  

  \subfigure[Carbon Monoxide (CO)]{\label{fig:PDF_x1_temp}
    \includegraphics[scale=\scaleFittings]{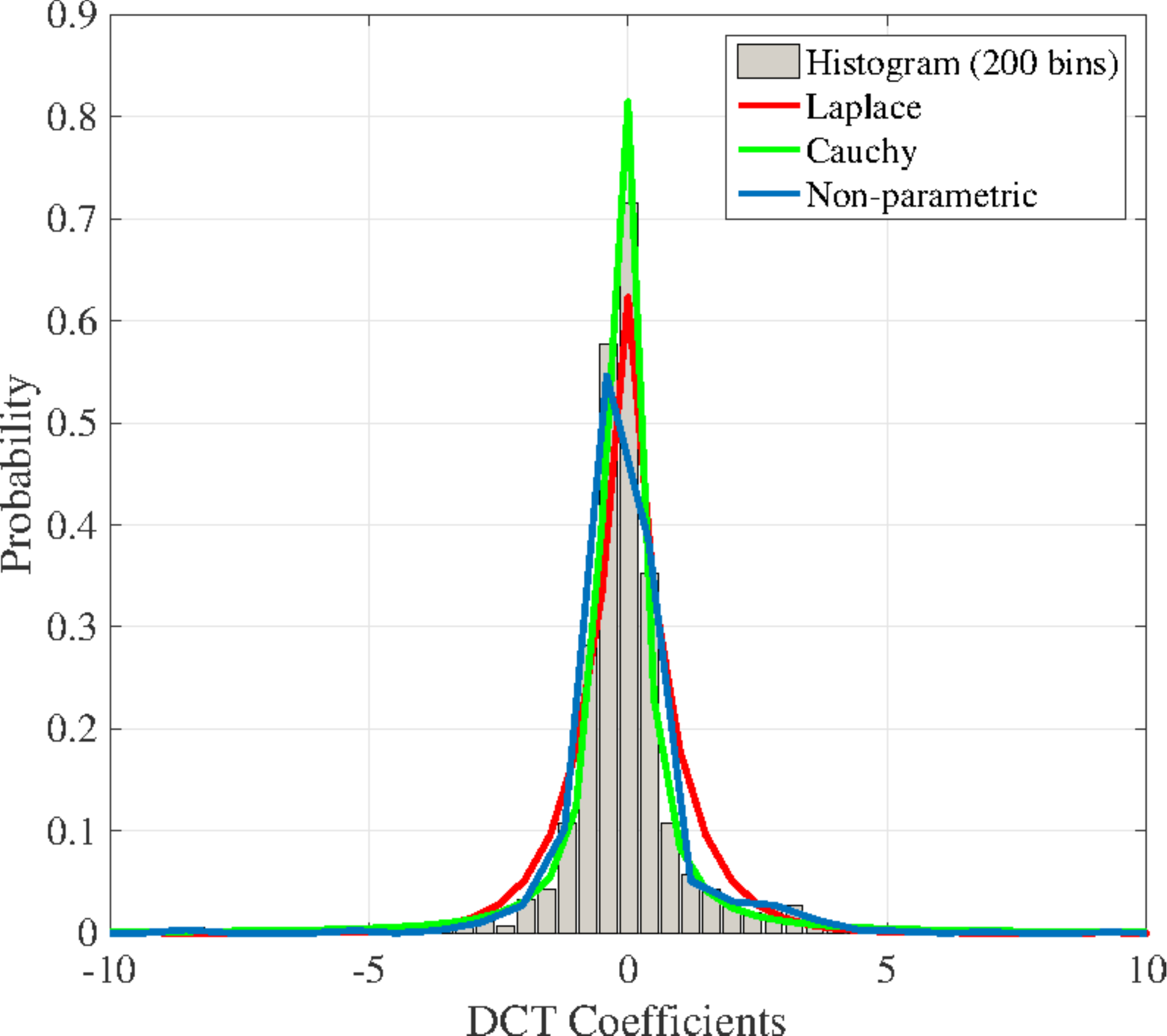} 
  }
  \hfill
  \subfigure[Nitrogen dioxide (NO$_2$)]{\label{fig:PDF_x2_hum}
    \includegraphics[scale=\scaleFittings]{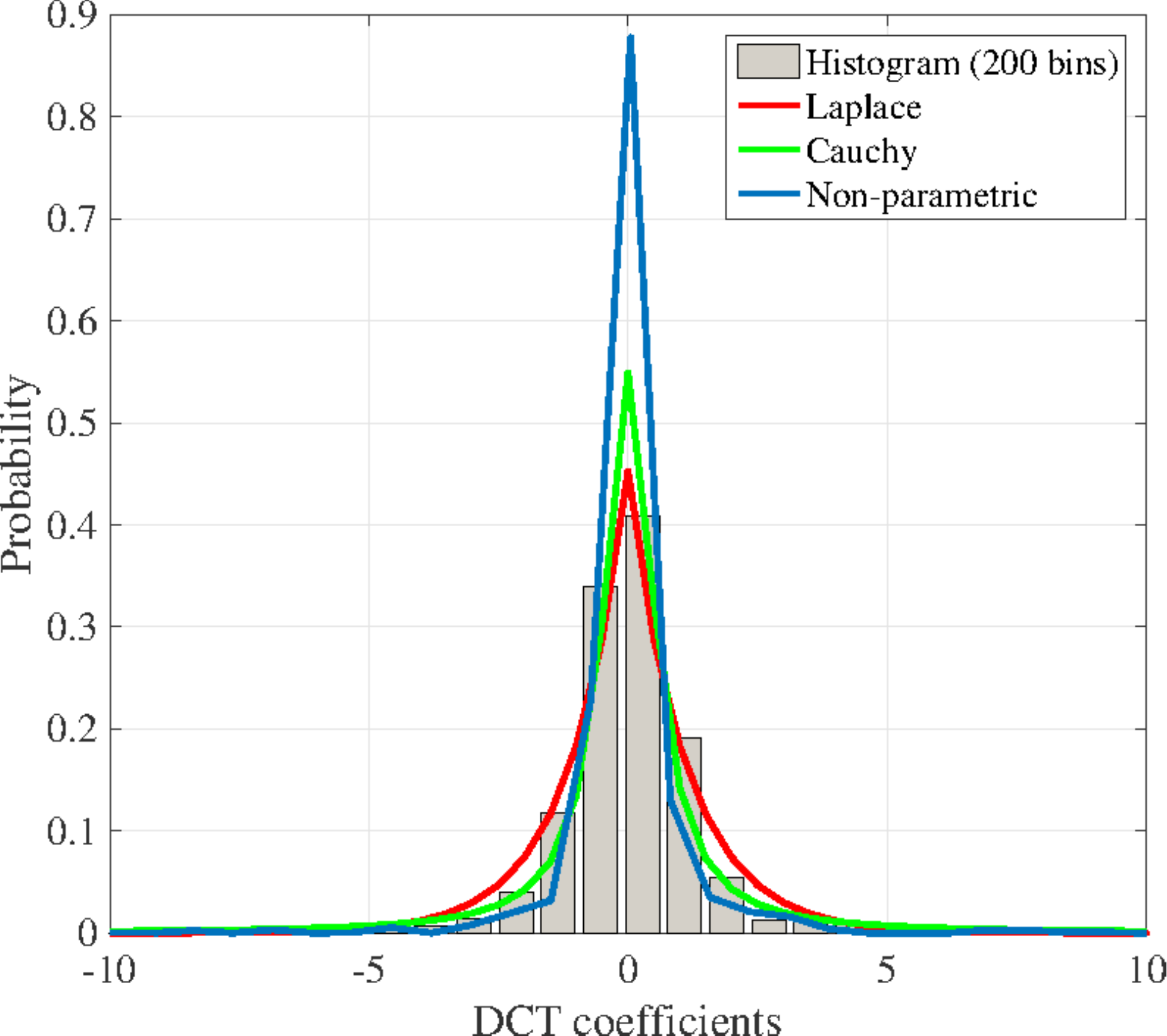}
  }
  \hfill
  \subfigure[Sulfur dioxide (SO$_2$)]{\label{fig:PDF_w1}
    \includegraphics[scale=\scaleFittings]{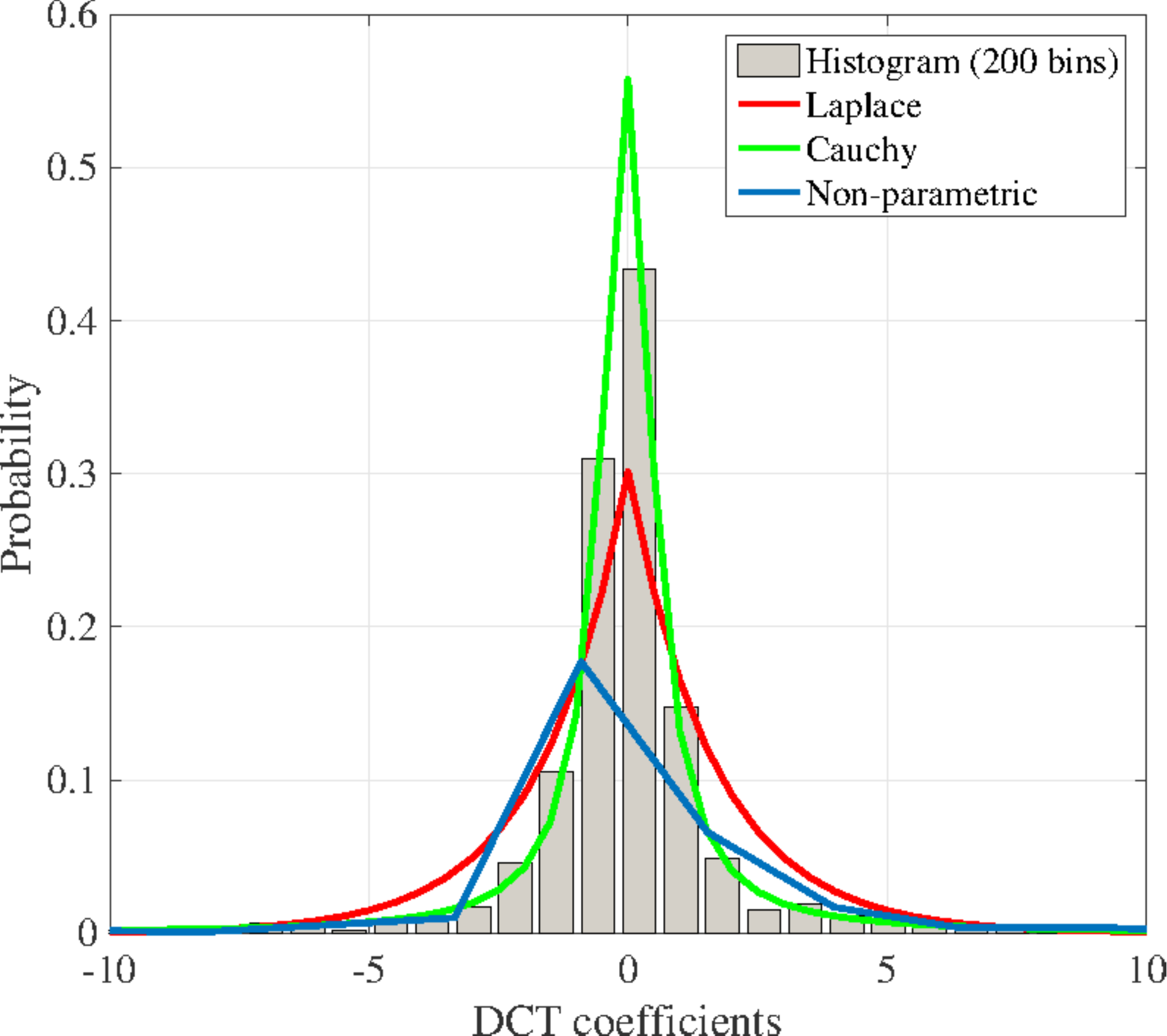}
  }
  \caption{
    Fitting of distributions (Laplace, Cauchy and KDE with a Gaussian
    kernel) on  DCT coefficients of different air pollutants in the
    EPA dataset~\cite{epa}. 
  }
  \label{fig:Fitting_Distr}
\end{figure*}

\subsubsection{Marginal Statistics and Copula Parameters}
\label{sec:StatisticsSelection}


To select the most appropriate marginal distribution for DCT coefficients of each~$\bm{s}^{(l)}$, with~$l =
1,2,3$, we performed fitting tests using the training set. The Laplace, the Cauchy, and the
non-parametric distribution---via KDE with a Gaussian Kernel---were fitted to the data using
the Kolmogorov-Smirnov test~\cite{massey1951kolmogorov}, with significance level set to~$5\%$. The
results, which were averaged over all the blocks in the training set, are reported in
Table~\ref{tab:KStest}  and Fig.~\ref{fig:Fitting_Distr}. We can observe that the Cauchy distribution
gives the best fit for the CO and SO$_2$ data, whereas the Laplace distribution best describes the
statistics of the NO$_2$ data. The parameters of the  distributions were estimated via maximum
likelihood estimation (MLE), resulting in $\hat{\beta}_{\text{CO}}=0.6511$,
$\hat{\beta}_{\text{SO}_2}=0.9476$ for the Cauchy distributions, and $\hat{b}_{\text{NO}_2}=2.3178$
for the Laplace distribution; recall the expressions for the pdf of these distributions
in~\eqref{Eq:LaplaceDistribution} and~\eqref{Eq:CauchyDistribution}. We also estimated the mean
values of the DCT coefficients, which were very close to zero for all
distributions.

We now elaborate on the estimation of the parameters of the different
copulas, described in Section~\ref{sec:CopulaFamilies}. Using
standard MLE \cite{bouye2000copulas}, we
calculate  the correlation matrix $\bm R_G$ for the Gaussian copula, the pairwise correlation
values of which are presented in Table \ref{tab:PairwiseEstimators}. Moreover,
we estimate the correlation matrix  $\bm R_t$  and the degrees-of-freedom parameter $\nu$ for the Student's $t$-copula
via approximate MLE \cite{bouye2000copulas}. The latter method fits a Student's $t$-copula by maximizing an objective function that approximates the profile
log-likelihood for the degrees-of-freedom parameter. For the ensemble of the three pollutants we find
the optimal value to be $\nu=89.91$,
whereas the values corresponding to each pair of pollutants are in Table \ref{tab:PairwiseEstimators}.
Table~\ref{tab:PairwiseEstimators} also reports the pair-wise maximum-likelihood estimates
\cite{genest1993statistical} of the $\xi$ parameter for different bivariate Archimedean copulas [cf.\
\eqref{Eq:ArchimedeanCopula}]. We consider bivariate Archimedean copulas for their
simplicity, i.e., they are parameterized by a single parameter. This, however, limits their modeling
capacity and makes them less accurate than, for example, Elliptical copulas \cite{nelsen2006introduction}.

\begin{table}[t]
\caption{Asymptotic $p$-values during the Kolmogorov-Smirnov fitting
tests to find the marginal distribution of the DCT coefficients of the data.}
\label{tab:KStest}
\centering 
\tabcolsep=0.1cm
\footnotesize
\begin{tabular}{|c|c|c|c|} 
\hline
        & Laplace   & Cauchy      & KDE \\ [0.15ex]\hline\hline
CO      & $0.0031$  & $0.6028$    & $9.4218\times10^{-20}$ \\ [0.1ex]
NO$_2$  & $0.5432$  & $0.1441$    & $2.2777\times10^{-21}$ \\ [0.1ex]
SO$_2$  & $0.0471$  & $0.9672$    & $1.0626\times10^{-21}$ \\ [0.1ex]\hline
\end{tabular}
\end{table}

\begin{table}[t]
\caption{Pairwise Copula Parameter Estimates.}
\label{tab:PairwiseEstimators}
\centering 
\footnotesize
\begin{tabular}{|c|c|c|c|} 
\hline
Parameters & $(\text{CO},\text{NO}_2)$   & $(\text{NO}_2,\text{SO}_2)$  &
$(\text{CO},\text{SO}_2)$\\\hline\hline 
Correlation & $0.7025$ & $0.8126$ & $0.8563$ \\
Degrees of Freedom, $\nu$  & $35.56$ & $35.56$ & $490.95$\\
$\xi$ (\text{Clayton}) & $1.5770$ & $2.3004$ & $2.7655$ \\
$\xi$ (\text{Frank}) & $6.6760$ & $8.8767$ & $11.0249$\\
$\xi$ (\text{Gumbel}) & $2.0877$& $2.5874$ & $3.1619$\\\hline
\end{tabular}
\end{table}

\subsubsection{Performance Evaluation of the Proposed Algorithm}
\label{subsec:ExpA}

\begin{figure}[t]
\centering
\subfigure[]{%
  \includegraphics[scale=0.43,trim=0.5cm .1cm 0.5cm 0.5cm]{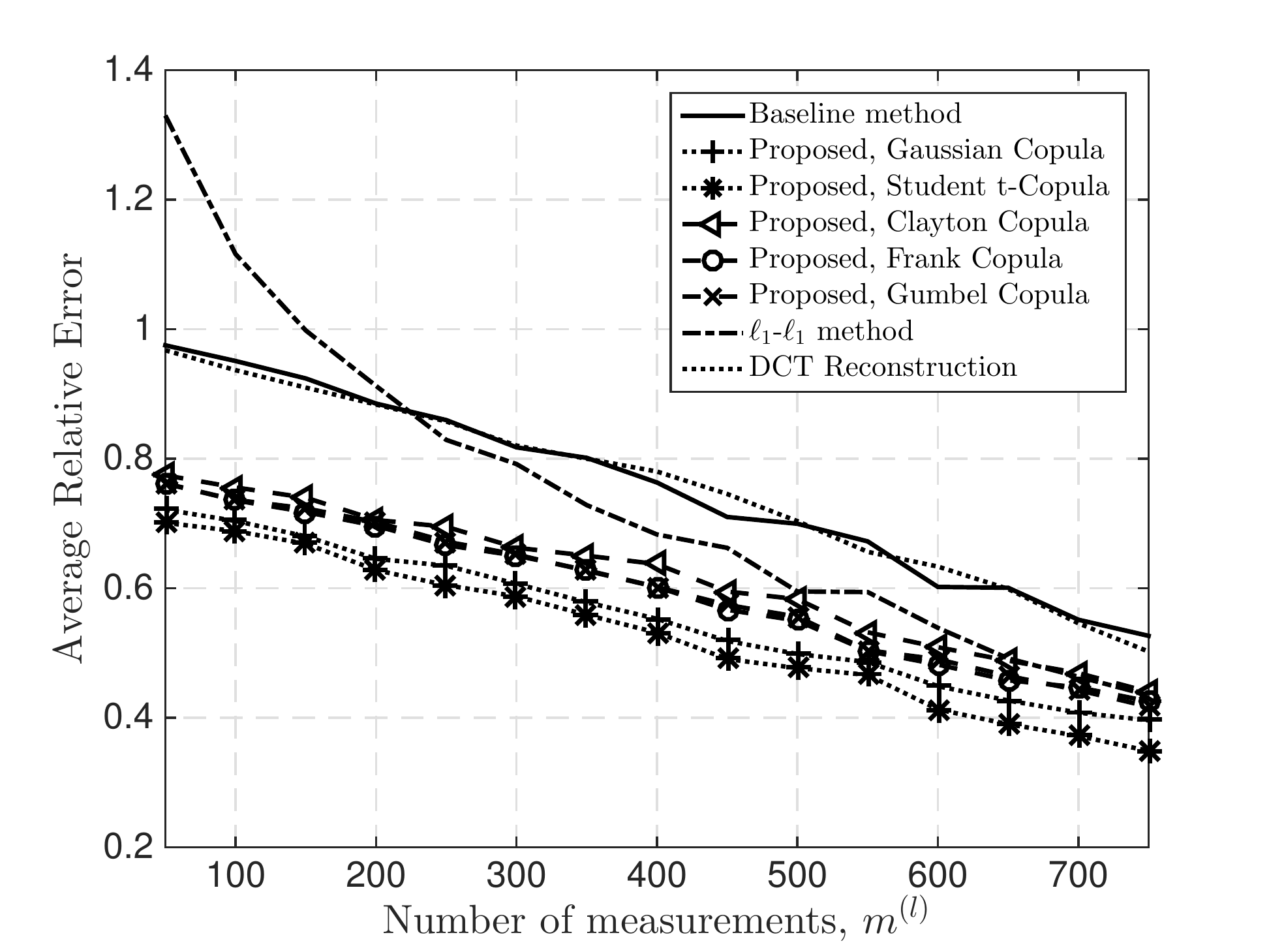} 
  \label{fig:single_SI}}
\quad
\subfigure[]{%
  \includegraphics[scale=0.43,trim=0.5cm .1cm 0.5cm 0.5cm]{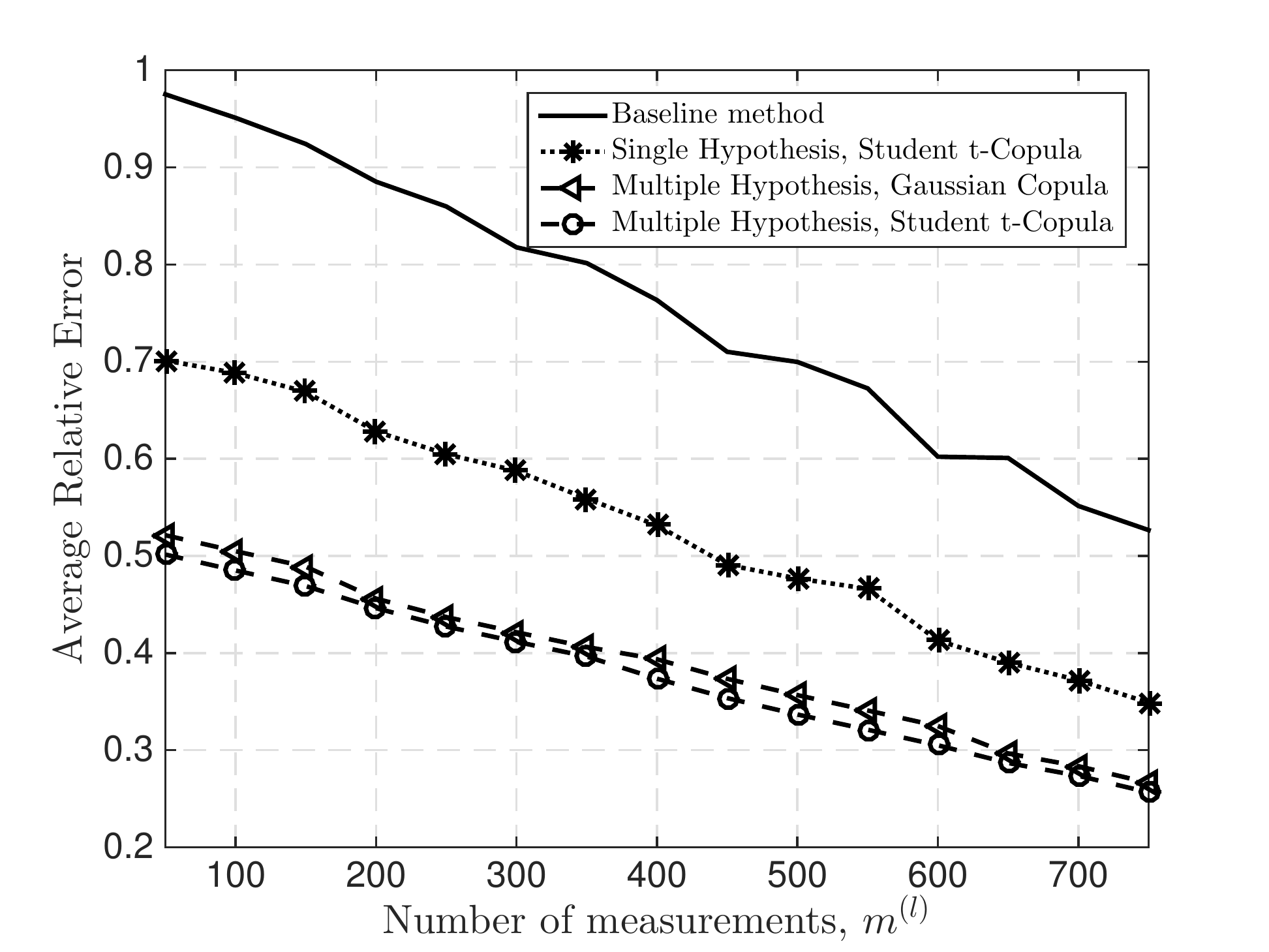}
  \label{fig:multihyp}} 
\caption{
  Reconstruction performance of CO signals using as side information (a)
only signals of NO$_2$, and (b) both signals of NO$_2$ and SO$_2$. The baseline method refers to the
no side information case, i.e., \cite{haupt2008compressed,luo2010efficient}.
} 
\label{fig:part1}
\end{figure}

\textcolor{black}{
We now describe how we evaluated the performance of our method against state-of-the-art reconstruction
algorithms. Simulating the data collection approach described in Section~\ref{sec:PropDataRec}, for every
vector of readings $\bm{x}^{(l)}$ in the test dataset, we
obtained its measurements as~$\bm{y}^{(l)}=\bm\Phi^{(l)}\bm{x}^{(l)}$. 
Similar to Section~\ref{sec:resultsSyntheticData}, we varied the number of measurements $m^{(l)}$ from $50$ to $750$ and, for each $m^{(l)}$, we
generated $50$ different matrices $\bm{\Phi}$ (independently). We will report the average [over the
$\bm{\Phi}$'s and over all the points $\bm{x}^{(l)}$ in the test dataset] relative
error~$\|\bm{x}^{(l)}-\widehat{\bm{x}}^{(l)}\|_2/\|\bm{x}^{(l)}\|_2$ as a function of $m^{(l)}$.
}

In the first set of experiments, we used the NO$_2$ data to aid the reconstruction of the CO
readings and considered the following methods: \textit{(i)} the proposed 
copula-based belief propagation algorithm, running for $50$ iterations and using five different
bivariate copulas for modelling the joint distribution, namely, the Gaussian, the Student's $t$, the
Clayton, the Frank, and the Gumbel copulas;
\textit{(ii)} the $\ell_1$-$\ell_1$ minimization method;\footnote{
  \textcolor{black}{
  The
  $\ell_1$-$\ell_1$ minimization problem~\eqref{eq:L1L1minDefinition} is solved using the code
  in~\cite{MotaGithubDocs}; a detailed explanation of the solver can be found therein. The
  experiments in~\cite{mota2014compressed,mota2014glob} show that such a solver finds medium-accuracy
  solutions to \eqref{eq:L1L1minDefinition} efficiently.
  }
}
\textit{(iii)} the baseline
method~\cite{haupt2008compressed,luo2010efficient}, which applies Bayesian CS~\cite{baron2010bayesian}
to recover the CO data independently; and, as a sanity check, \textit{(iv)} simply keeping the
$m^{(l)}$ largest (in absolute value) DCT coefficients. 

Fig.~\ref{fig:single_SI}
depicts the \textcolor{black}{relative reconstruction error}
versus the number of measurements $m^{(l)}$. It is clear that the proposed
algorithm and \textcolor{black}{$\ell_1$-$\ell_1$ minimization} 
efficiently exploited the side information and were able to improve the performance with respect to the baseline
method~\cite{haupt2008compressed,luo2010efficient}. 
\textcolor{black}{
When the number of measurements was small ($<200$), 
the baseline method outperformed $\ell_1$-$\ell_1$ minimization; 
this is because, with few measurements, the side information was actually hindering reconstruction;
recall that $\ell_1$-$\ell_1$ minimization assumes the side information to be of the same kind as the
signal to reconstruct.
}
Furthermore, it is clear that the proposed algorithm
systematically outperformed \textcolor{black}{$\ell_1$-$\ell_1$ minimization} \cite{mota2014compressed}
for all the considered copula functions. The best performance was achieved by the Student $t$-copula
function, providing average relative error reductions of up to $47.3\%$ compared to
$\ell_1$-$\ell_1$ minimization. 
\textcolor{black}{
We mention that, contrary to most results in
compressed sensing, the results of Fig.~\ref{fig:single_SI} fail to exhibit a precise phase transition.
This is because the representation of the data is not exactly sparse, only compressible. That can be
seen in the plot, as the baseline method~\cite{haupt2008compressed,luo2010efficient} had a very similar
performance to the DCT reconstruction, i.e., keeping only the largest DCT coefficients. This also shows
that, in this case, what allowed both our method and $\ell_1$-$\ell_1$ minimization to achieve better
performance was the proper use of the (correlated) side information. 
}

In another experiment, we reconstructed CO readings using as side information data from the other two
pollutants, i.e.,  NO$_2$ and SO$_2$. Fig. \ref{fig:multihyp} shows the \textcolor{black}{average
  relative error} of the proposed algorithm with one and two side information signals, and also the
baseline method~\cite{haupt2008compressed,luo2010efficient}. It is clear that the more side
information signals there are, the better the performance of our algorithm. We also observe that the
Student's $t$-copula lead to a performance better than the Gaussian copula; this was because the former
depends on more parameters than the latter, giving it a larger modeling
capacity~\cite{breymann2003dependence}.

\subsubsection{Evaluation of the Aggregated System Performance}
\label{sec:AggSystemPerf}

\begin{figure}[t]
\centering
\subfigure[]{%
  \includegraphics[scale=0.45,trim=0.5cm .1cm 0.5cm 0.5cm]{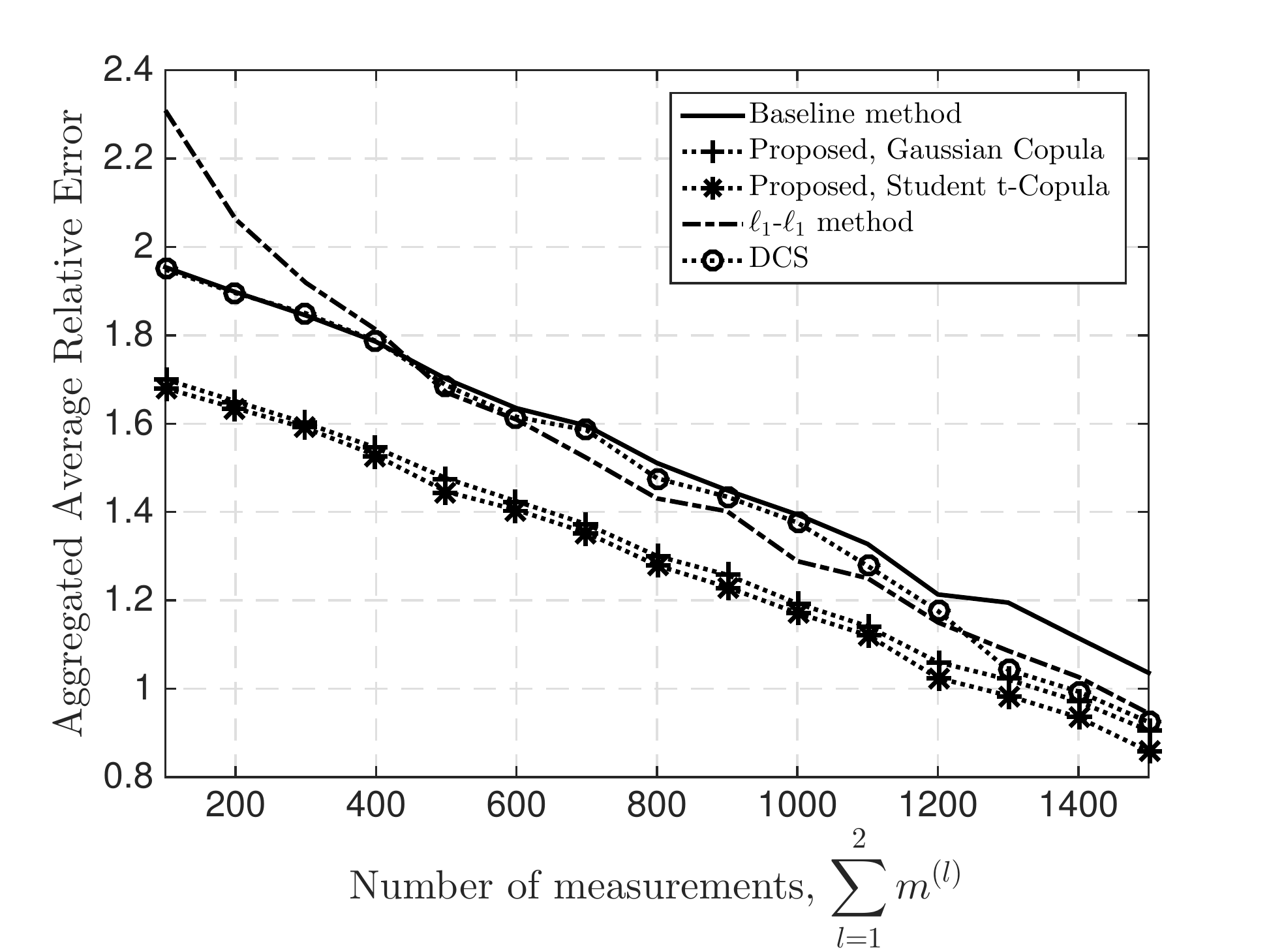}
  \label{fig:two_pol}}
\quad
\subfigure[]{%
  \includegraphics[scale=0.45,trim=0.5cm .1cm 0.5cm 0cm]{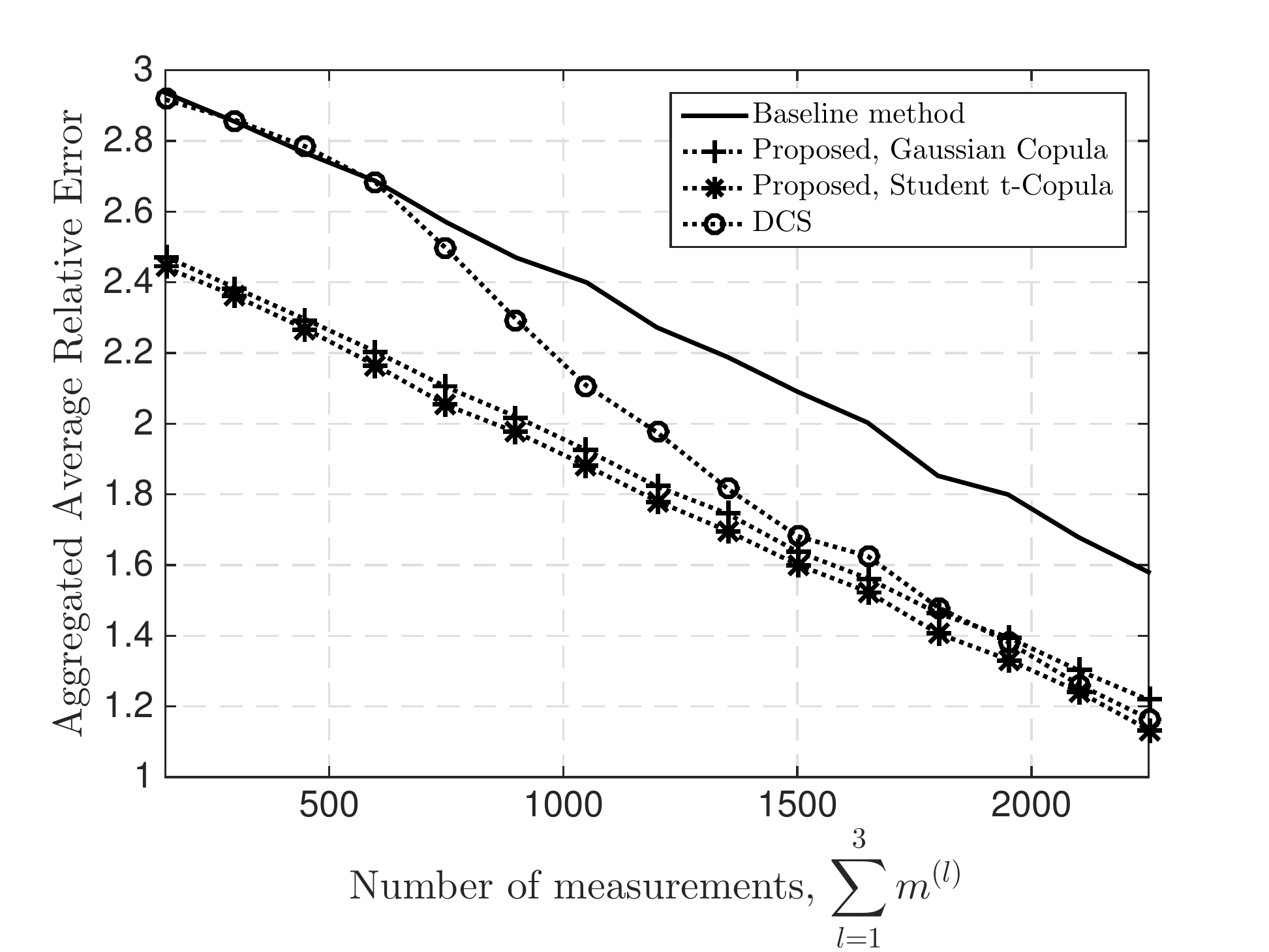}
  \label{fig:three_pol}} 
\caption{Performance comparison of the proposed successive reconstruction architecture with the DCS, the ADMM-based and the baseline systems when we assume (a)
two air pollutants (CO and NO$_2$), and (b) three air pollutants (CO, NO$_2$ and SO$_2$). }
\label{fig:part2}
\end{figure}

We now describe the experiments conducted to evaluate the sequential reconstruction
algorithm in which the readings are reconstructed consecutively.
First, we focus on the scenario where two pollutants are measured, and
we compare the following schemes: \textit{(i)} the proposed sequential scheme, using the Gaussian and
the Student $t$-copula models (as shown in Section \ref{subsec:ExpA}), they perform better than other
copulas); \textit{(ii)} sequential data recovery using \textcolor{black}{$\ell_1$-$\ell_1$
minimization} \cite{mota2014compressed}; 
\textit{(iii)} the DCS setup\footnote{In the classical DCS scenario, each signal of interest is
  constructed by many readings of the same sensor. In order to have a fair comparison with our
  design, we have modified this framework by assuming that each signal of interest contains readings
  from different sensors observing the same source. \textcolor{black}{In our experiments we used
    $\omega_1=\dots=\omega_\ell=1$.}}~\cite{duarte2005distributed}; and \textit{(iv)} the baseline
system in which each source is independently reconstructed using Bayesian
CS\cite{baron2010bayesian}.

The performance metric is expressed as the \textcolor{black}{aggregated average relative error for all 
signals,
$\sum_{l=1}^{\ell}\|\bm{x}^{(l)}-\widehat{\bm{x}}^{(l)}\|_2/\|\bm{x}^{(l)}\|_2$,
versus the total number of measurements $\sum_{l=1}^{\ell}m^{(l)}$.} 
Fig.~\ref{fig:two_pol} shows that the systems based on \textcolor{black}{$\ell_1$-$\ell_1$
minimization} and on DCS leverage both the inter- and intra-source dependencies between the pollutants, resulting
in an improved performance with respect to the baseline system. However, when the number of
measurements is small ($\sum_{l=1}^{2}m^{(l)}<400$), we see that \textcolor{black}{$\ell_1$-$\ell_1$
minimization} performs poorly compared to the other methods. The proposed system with the Student
$t$-copula model systematically outperforms all the other schemes, bringing \textcolor{black}{aggregated average relative error} improvements of up to $27.2\%$ and $13.8\%$ against \textcolor{black}{$\ell_1$-$\ell_1$
minimization} and DCS, respectively.

When three pollutants are measured, we compared all the previous schemes, except the one based on
\textcolor{black}{$\ell_1$-$\ell_1$ minimization}, since it does not handle multiple side information
signals. Fig.~\ref{fig:three_pol} shows that DCS delivers superior performance compared to the
baseline system, which is more noticeable when $\sum_{l=1}^{3}m^{(l)}>600$. Furthermore,  the proposed design with the Student $t$-copula model provides significant \textcolor{black}{aggregated average relative error} reductions
of up to $19.3\%$ when compared to DCS~\cite{duarte2005distributed}. It is important to notice that the proposed design significantly outperforms the other schemes when the number of measurements is small. 

\begin{table*}[t]
\textcolor{black}{\caption{Number of Measurements and Energy Consumption at the Wireless Nodes for Two Different Data Recovery Quality Levels. Two Pollutants (CO, NO$_2$) are Measured.}} 
\label{tab:EnergyConsumTwoSources}
\centering 
\tabcolsep=0.1cm
\begin{tabular}{|c|c|c|c|||c|c|c|} 
\hline
& \multicolumn{3}{|c|||}{\text{Medium Data Recovery Quality}} & \multicolumn{3}{|c|}{\text{High Data Recovery Quality}}\\[0.1ex]
\hline
 &  Baseline & $\ell_1$-$\ell_1$ & Proposed &  Baseline & $\ell_1$-$\ell_1$ & Proposed \\[0.1ex]
\hline\hline
Aggregated average relative error &   1.4046 &  1.4230 &  1.3957 & 1.1148 &  1.1044 &  1.0969\\
Total number of measurements & 950 & 850 & 550 & 1400 & 1300 & 1050\\
$E^{\text{HW}}_{\text{proc.}} \:(\text{J})$  & $4.84\times10^{-6}$ & $4.33\times10^{-6}$ & $2.80\times10^{-6}$ & $7.14\times10^{-6}$ & $6.63\times10^{-6}$ & $5.35\times10^{-6}$\\
$E^{\text{SW}}_{\text{proc.}} \:(\text{J})$ & $46.51\times10^{-6}$ & $41.62\times10^{-6}$ & $26.93\times10^{-6}$ & $68.54\times10^{-6}$ & $63.64\times10^{-6}$ & $51.41\times10^{-6}$\\
$E_{\text{Tx}} \:(\text{J})$ & $0.8586$ & $ 0.7668$ & $0.4968$ & $1.2636$ & $1.1718$ & $0.9450$\\
$E_{\text{total}}\:(\text{J})$ & $0.8586$ & $0.7668$ & $0.4968$ & $1.2636$ & $1.1718$ & $0.9450$ \\
\hline 
\end{tabular}
\end{table*}

\subsubsection{Evaluation of the System Performance under Noise}

\begin{figure}[t]
\centering
\subfigure[]{%
  \includegraphics[scale=0.45,trim=0.5cm .1cm 0.5cm 0.5cm]{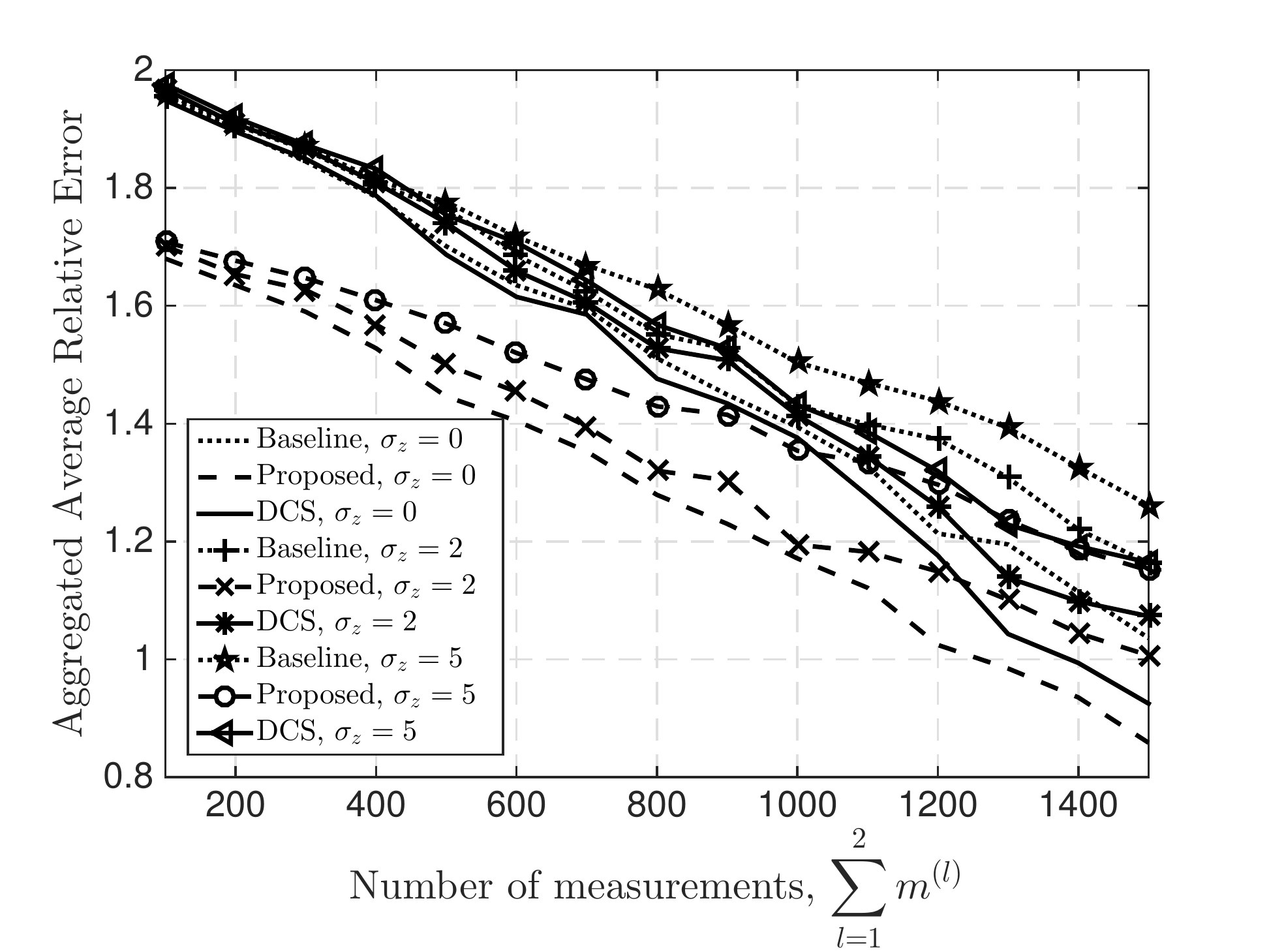}
  \label{fig:two_pol_noise}}
\quad
\subfigure[]{%
  \includegraphics[scale=0.45,trim=0.5cm .1cm 0.5cm 0cm]{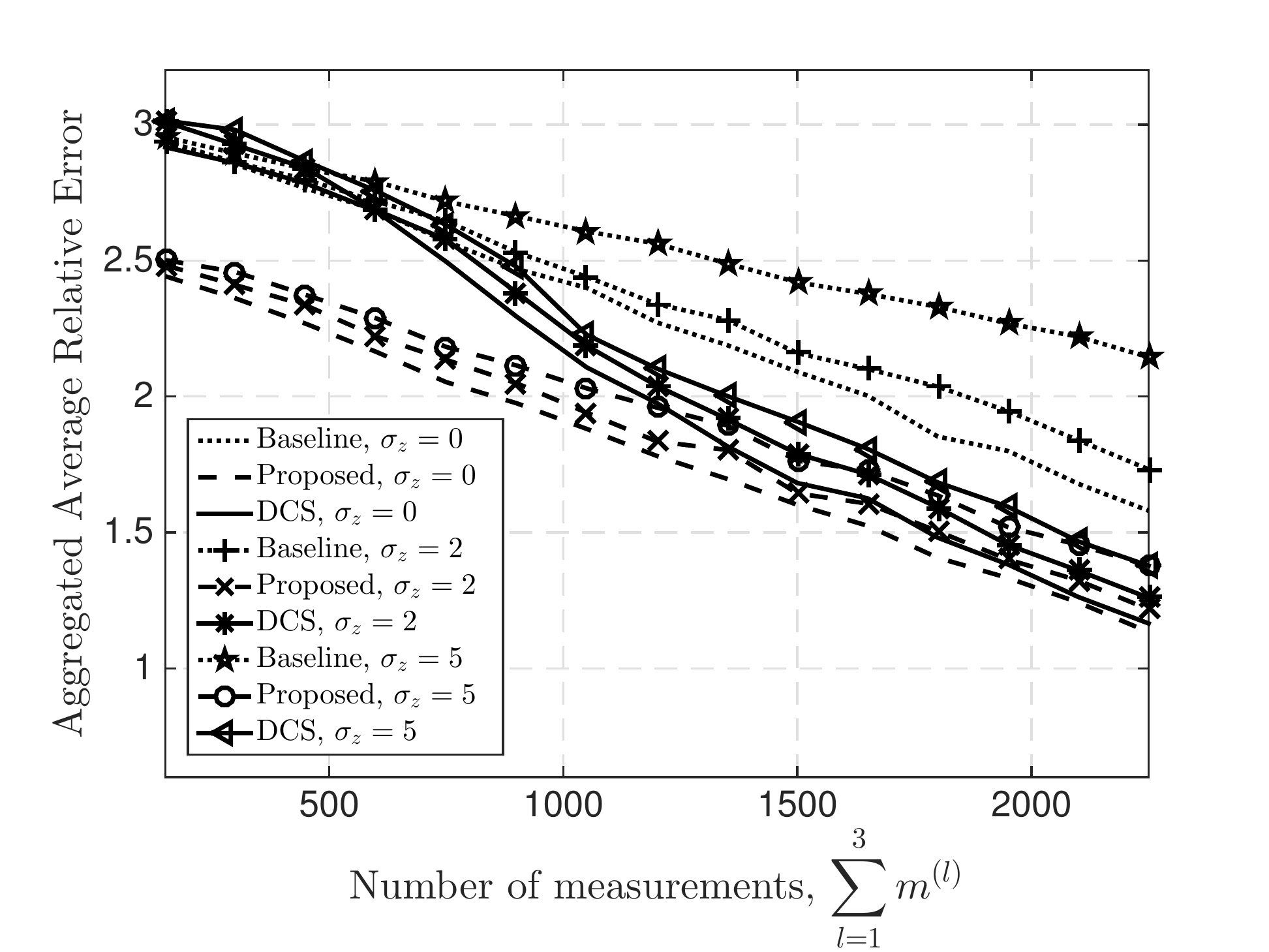}
  \label{fig:three_pol_noise}} 
\caption{Performance comparison of the proposed system with the DCS setup
when we assume different noise levels ($\sigma_z = 0,2,5$) for (a)
two sources (CO and NO$_2$), and (b) three sources (CO, NO$_2$ and SO$_2$).}
\label{fig:part3}
\end{figure}

We evaluate the robustness of the proposed successive data recovery architecture against
\textcolor{black}{imperfections in the communication medium}. As explained in
Section~\ref{sec:PropDataRec}, \textcolor{black}{we model such imperfections using a zero-mean white
  Gaussian noise
  component~$\bm{z}^{(l)}\sim\mathcal{N}(\bm{0},\sigma_z\bm{I})$
additive to the measurements, where~$\sigma_z$ is the noise standard deviation\footnote{We assume that the standard deviation of the noise is the same for all sources; hence, we drop the superscript $(l)$.} and~$\bm{I}$ is the~$m^{(l)}\times m^{(l)}$ identity matrix.} In this experiment, we vary the noise level as
$\sigma_z \in \{0,2,5\}$ and calculate the \textcolor{black}{aggregated average relative error}
as a function of the total number of measurements. 
 We first consider the case in which two pollutants are gathered by each device. The considered
 schemes are (\textit{i}) the proposed system with successive
data recovery using the copula-based algorithm (the Student's $t$ copula is used);  (\textit{ii})  DCS~\cite{duarte2005distributed}, and; (\textit{iii}) the baseline system~\cite{haupt2008compressed,luo2010efficient}. Figs. \ref{fig:two_pol_noise} and
\ref{fig:three_pol_noise} show that the proposed system delivers superior performance compared to the competing systems for moderate ($\sigma_z=2$) and high ($\sigma_z=5$)
noise. Moreover, we observe that the proposed algorithm is
robust against noise, especially, when the number of measurements is small. In particular, the \textcolor{black}{aggregated
average relative error}
increases on average $5.8\%$ ($\sigma_z=2$) and $18.7\%$ ($\sigma_z=5$) with respect to the noiseless case. 

In case three pollutants are measured, the proposed system systematically outperforms the DCS scheme
and the baseline system, under both moderate and high noise. Moreover, the proposed design continues
to demonstrate robustness against noise, with the \textcolor{black}{aggregated average relative error}
 increasing on average only $4.2\%$ ($\sigma_z=2$) and $10.1\%$ ($\sigma_z=5$) compared to the
 noiseless case. It is clear that the robustness of the proposed system increases with the number of
 pollutants.

\textcolor{black}
{
\subsection{Energy Consumption Analysis}
\label{sec:EnergyConsumption}
We now study the impact of the proposed system on reducing the number of
measurements and in turn the energy consumption of the wireless nodes, for
a given data reconstruction quality. The energy consumption at each node
is broken down into a sensing, processing and transmission part:
$E_\text{total} = E_\text{sens.} + E_\text{proc.} + E_\text{Tx}$~\cite{landsiedel2005accurate}. 
The sensing part, $E_\text{sens.}$, depends on the amount of censored data; hence, its energy consumption is the same for the proposed and the baseline system. We thus focus our comparison on the energy consumption due to the processing and transmission parts. Following a typical IoT design, we assume that the nodes are equipped
with the MSP430 micro-controller~\cite{instruments1999msp430} and that communication
adheres to LoRa~\cite{lorawan}.  MSP430 architectures~\cite{instruments1999msp430}
are typically built around a 16-bit CPU running at 25 MHz, with a voltage
  supply of $V = 1.8-3.6$~Volt and a current of $I=200$ $\mu$A/MIPS in the
active mode. As discussed in Section~\ref{sec:PropDataRec},  every node 
generates a pseudorandom number,
computes the product between this number and the censored value, adds it
to the sum of the previous relayed values, and sends the final value to the
next node. This operation is repeated per  measurement and  source~$l\in\{1,...,\ell\}$.
Neglecting the pseudorandom number generation part, the encoding operation
boils down to a multiply-and-accumulate (MAC) operation. The MSP430 CPU\
cycles needed for a single signed 16-bit MAC operation are 17 for a hardware implementation or between $143$ and $177$ for a software implementation~\cite{instruments1999msp430}.
Therefore, the time to perform a single MAC operation\footnote{We consider
the higher value on the number of cycles for software.} is $t^{\text{HW}}=\frac{17\:\text{cycles}}{25\:\text{MHz}}=0.68\:\mu{s}$
or $t^{\text{SW}}=\frac{177\:\text{cycles}}{25\:\text{MHz}}=7.08\:\mu{s}$.
The total time to encode the measurements at each device can then be calculated
as $t^{\text{(\textbullet)}}_\text{proc.}=\sum_{l=1}^{\ell}m^{(l)}\times
t^{\text{(\textbullet)}}$ and the total processing energy as $E^{\text{(\textbullet)}}_{\text{proc.}}=V\cdot
I\cdot t^{\text{(\textbullet)}}_\text{proc.}$, where $\text{(\textbullet)}$
stands for HW\ or SW.
In order to calculate the energy consumption for transmission, we used the
LoRa energy consumption calculator from Semtech~\cite{semReport,semWeb}.
For a typical 12-byte payload packet with a 14 dBm power level, a current
at 44 mA and a spreading factor of 7, the transmission energy consumption
was estimated at 5.4 mJ. In the scenario where two pollutants (CO and NO$_2$) are encoded [and no noise is assumed in the communication medium], Table~\ref{tab:EnergyConsumTwoSources}
reports the number of measurements and the energy consumption at the nodes
for the baseline system, the system using $\ell_1$-$\ell_1$ minimization
\cite{mota2014compressed}, and the proposed system. It is worth observing
that  the processing energy  is negligible compared to the energy consumed
by the transceiver.  It is  evident that for a comparable aggregated average
relative error the proposed system leads to a significant reduction in the
number of transmitted measurements compared to the competition, which translates
to critical energy savings at the nodes.
}

\section{Conclusion \textcolor{black}{and Future Work}}
\label{sec:conclusion}
We addressed the problem of \textcolor{black}{data recovery from compressive measurements} in
large-scale WSN applications, such as air-pollution monitoring. In order to efficiently capture
statistical dependencies among heterogeneous sensor data, we used copula functions
\cite{sklar1959fonctions,nelsen2006introduction}. This enabled us to devise a novel CS-based
reconstruction algorithm, built upon belief propagation
\cite{mackay2003information,cowell2006probabilistic},   which leverages multiple heterogeneous
signals (e.g., air pollutants) as side information in order to improve reconstruction. Experiments
using \textcolor{black}{synthetic data and real sensor data from} the USA EPA showed that the proposed scheme significantly improves the quality of
data reconstruction with respect to prior state-of-the-art methods
\cite{haupt2006signal,baron2005information,mota2014compressed}, even under sensing and communication
noise. \textcolor{black}{Furthermore, we showed that, for a given data reconstruction quality, the proposed scheme offers low encoding complexity and reduced radio transmissions compared to the state of the art, thereby leading to energy savings at the wireless devices.} We conclude that our design
effectively meets the demands of a large-scale monitoring application. 
\textcolor{black}{Future work should concentrate on assessing the method on alternative datasets, such as the Intel-Berkeley Lab dataset~\cite{intellabWSN}, the dataset from the Center for Climatic Research~\cite{CCR_dataset}, and the indoor dataset from the University of Padova~\cite{crepaldi2007design}.}   


%

%
%
%
%
%

\ifCLASSOPTIONcaptionsoff
  \newpage
\fi



\bibliographystyle{IEEEtran}
\bibliography{bare_adv}
\end{document}